\documentclass[aps,preprintnumbers,amsmath,amssymb,nofootinbib,eqsecnum, preprintnumbers]{revtex4}
\usepackage{eurosym}
\usepackage{amsfonts}
\usepackage{amsmath}
\usepackage{amssymb,epsf}
\usepackage{color}
\usepackage{graphicx}
\usepackage{natbib}
\usepackage{float}
\usepackage{caption}
\usepackage{subfig}
\usepackage{epstopdf}

\begin{document}
	\title{Optical Signatures of Einstein-Euler-Heisenberg AdS/dS  Black Holes in the light of Event Horizon Telescope}
\author{Khadije Jafarzade$^{1}$\footnote{
email address: khadije.jafarzade@gmail.com}, Zeynab Bazyar$^{2}$\footnote{
email address: hzp.bazyar@gmail.com}, Sara Saghafi$^{1}$\footnote{
email address: s.saghafi@umz.ac.ir}, Kourosh Nozari$^{1}$\footnote{
email address: knozari@umz.ac.ir}}
\affiliation{$^{1}$Department of Theoretical Physics, Faculty of Science, University of Mazandaran,
P. O. Box 47416-95447, Babolsar, Iran\\
$^{2}$Department of Physics, Isfahan University of Technology, 84156-83111,
Isfahan, Iran}	
	\begin{abstract}
Recent observations of the supermassive black holes $ M87^{*} $ and Sgr A$^{*}$ by the Event Horizon Telescope (EHT) have sparked intensified interest in studying the optical appearance of black holes (BHs). Inspired by this, we carry out a study on the optical features of Einstein-Euler-Heisenberg-Anti de Sitter/de Sitter (EEH-AdS/dS) BHs, including the trajectories of photons, shadow geometrical shape, energy emission rate, and deflection of light in this spacetime.	Since, due to the nonlinear electrodynamics effects, photons propagate along null geodesics in an effective metric rather than the background metric, we first derive the effective metric of the EEH-AdS/dS BH. Then we study the null geodesics of the resulting effective metric and eventually compute the size of the EEH-AdS/dS BH shadow. To validate our results, we confront our results with the extracted information from EHT data of the supermassive BHs $ M87^{*} $ and estimate lower bounds for the shadow radius.

	\end{abstract}
	
	\maketitle
	
	\section{Introduction}
		
Recently, the EHT revealed the first stunning image of the supermassive BH at the center of the galaxy $ M87^{*}$, confirming the existence of BHs in the universe as the most interesting prediction of general relativity (GR) \cite{EventHorizonTelescope:2019dse,EventHorizonTelescope:2019uob,EventHorizonTelescope:2019jan,EventHorizonTelescope:2019ths,EventHorizonTelescope:2019pgp,EventHorizonTelescope:2019ggy}. Recent observations of EHT have opened new avenues for testing gravity theories through BH shadow observables. EHT observations probe the geometric (metric) structure of spacetime without assuming a specific theory of gravity, making it an essential and powerful tool for testing predictions from different theories of gravity \cite{Psaltis:51}.  In other words, these observations offer a means to distinguish between GR and modified gravity theories while providing insights into the astrophysical properties of the observed BHs. The experimental results reported by EHT have enhanced our insight and understanding of GR that is not available from other astrophysical experiments and systems. Additionally, the horizon-scale images of the supermassive BHs observed by the EHT probe gravitational fields vastly differ from those probed in all other GR tests, with or without black holes \cite{Psaltis:51}.

Simulated images of BHs surrounded by optically thin emitting material often display two distinct features: a central dark region and a narrow, bright ring known as the photon ring. This ring results from extreme gravitational lensing and appears superimposed on the more direct emission. It traces a theoretical path on the image plane that corresponds to light rays orbiting near the BH along unstable photon trajectories. The shape and size of this so-called critical curve depend entirely on the spacetime geometry around the BH. In contrast, the characteristics of the central dim region, its size, shape, and depth, are influenced by the nature of the emitting matter. For instance, simulations based on spherical accretion models produce a prominent dark area referred to as the BH shadow, which fills the space enclosed by the photon ring. However, models involving equatorial accretion disks that reach down to the event horizon result in a smaller dark feature known as the inner shadow \cite{Dokuchaev:583,Chael:918}. This inner region is defined by the lensed outline of the equatorial event horizon and consists of photon paths that fall directly into the BH without intersecting the equatorial plane. While the photon ring and critical curve are typically circular and relatively unaffected by the observer’s line of sight, the inner shadow is more sensitive to the viewing angle. It appears circular only when observed face-on and changes shape and position with inclination. Though current observations from the EHT do not yet have the resolution or dynamic range to detect the inner shadow, it remains a significant theoretical feature that future, more detailed imaging may uncover.

The formation of a BH shadow is strongly influenced by the angular momentum of photons. Its size and shape are determined by several factors, including the BH mass, spin (angular momentum), electric charge \cite{Vries:1999123}, the surrounding spacetime geometry \cite{Johannsen:446,Cunha:024039}, and the observer’s viewing angle. The image of a BH provides us with substantial information about jets and the dynamics of matter around BHs. Additionally, BH shadow can be used to extract information about distortions in the geometry of spacetime \cite{Pedro:5042,Wei:08030,Belhaj:215004}. The history of the study of the BH shadow began with the seminal paper of Synge \cite{Synge:1966okc},  who calculated the angular radius of the Schwarzschild BH shadow and showed that the boundary of the shadow is a perfect circle for a spherically symmetric BH. Subsequently, Bardeen analyzed the shadow of a Kerr BH  based on Carter's work \cite{Carter:1968rr} and argued that a rotating BH has an elongated shape in the direction of the rotation axis due to the dragging effect \cite{Bardeen:1972fi}. Afterward, the BH shadow became a fascinating topic in astrophysics and attracted much attention in a wide range of research in BH physics, such as BHs in modified GR 
\cite{Liu:858,Wang:05,Fang:08,Vagnozzi:40,Khodadi:26932,Bambi:100,Jafarzade:1a,Jafarzade:1b,Zheng:1d,Gao:1d,Erices:1a}, BHs surrounded by plasma \cite{Feng:73,Kumar:44,Maqsood:047,Atamurotov:2023,Briozzo:2023}, and BHs with nonlinear electrodynamics \cite{Bakopoulos:110,Zhong:103,Zhong:104,Aliyan:1b,Zeng:764,Guzman:11,Lambiase:48,Hamil:73}.

One of the enduring mysteries of GR is the nature of the singularity that remains hidden in BHs. The existence of a singularity in Maxwell's equations of electrodynamics causes the ultraviolet divergent self-energy of a point-like charge in classical dynamics. To solve the problem of infinite self-energy of a point charge, Born and Infeld generalized Maxwell electrodynamics \cite{Born:1934gh} and proposed nonlinear electrodynamics (NLED). Thereafter,  Euler and  Heisenberg proposed a new solution for the infinite self-energy of a charged particle, known as Euler-Heisenberg (EH) electrodynamics \cite{Heisenberg:1936nmg}.  This concept was developed in research and later played a crucial role in studying string theory \cite{Magnea:2004ai,Sciuto:2005sq}. The EH theory is one of the profound theories of nonlinear electrodynamics that, within a theoretical framework, describes the behavior of the electromagnetic field in the presence of electric and magnetic fields, while also quantitatively calculating the effects of quantum electrodynamics. Noticeably, as the electric and magnetic fields approach the critical limits of $E_{cr}\approx 10^{18}$ and $B_{cr}\approx 10^{10}$ \cite{Novello:1999pg}, quantum electrodynamic effects become particularly noticeable, including light-to-light scattering \cite{ATLAS:2017fur,ATLAS:2019azn} and vacuum birefringence \cite{Capparelli:2017mlv}. The EH Lagrangian describes the nonlinear interactions of electromagnetic fields due to closed electron loops \cite{Schwinger:1951nm}. For many years, it was believed that electromagnetic processes could be described using the EH Lagrangian, including photon-photon scattering \cite{Euler:1935zz,Euler:1935qgl} and photon splitting in the presence of a strong magnetic field \cite{Adler:1970gg,adler1}.

Nonlinear electrodynamics has also been influential in studying the BH shadow, which has been the subject of much astronomical exploration \cite{Okyay:2021nnh}. The motion of light under non-specific vacuum conditions can be considered as electromagnetic waves propagating through a classically dispersive medium \cite{Dittrich:1998fy}. In such a nonlinear framework, photons no longer move through null geodesics in Minkowski spacetime but instead move in effective geometry \cite{Plebanski:1970zz}. In 2000, Novello et al investigated the geometric aspects of light propagation in nonlinear electrodynamics. They used the EH nonlinear electrodynamics in the spacetime of regular BHs to understand the effective geometry \cite{Novello:1999pg}. Recently, due to the advancement of the nonlinear quantum electrodynamics, huge attention has been devoted to study of light propagation, including the geometry of light propagation in nonlinear electrodynamics \cite{Theodosopoulos:84,Allahyari:1a,Zeng:2022pvb} and the generalized Born-Infeld electrodynamics \cite{Kim:2022xum}.

The  Einstein-Euler-Heisenberg (EEH) model is an exact solution to the coupling of EH theory and GR whose solutions have been studied from various perspectives such as thermodynamics \cite{Cheng:yh}, thermal fluctuations \cite{Gursel:hm}, gravitational lensing \cite{Qiao:su}, accretion disks \cite{Jiang:yh}, shadow and quasinormal modes \cite{Lambiase:gd}. As an example in \cite{Myung:cv}, an analysis on the shadow of charged EEH BHs was presented by studying null geodesics in the background geometry, and theoretical predictions were confronted with the observational data of Sgr A$^{*} $. The results showed that the shadow diameter of these BHs is consistent with the EHT data in narrow ranges of electric charge. However, in Ref. \cite{Guzman:an}, the EEH BH shadow was investigated by studying the null geodesics of  the effective metric and was compared with shadow observations reported by EHT for both supermassive BHs Sgr A$^{*} $ and $ M87^{*} $. In \cite{Gursel:hm}, the authors considered EEH BHs and studied thermal fluctuations from two different perspectives. They found that the effect of vacuum fluctuations becomes apparent for large values of the electric charge and EH parameter. In other words, as nonlinear effects become dominant, it becomes relatively easier to make further comments on the corrections due to vacuum polarization. A study on the phase structure of EEH-AdS BHs in both canonical and grand canonical ensembles was presented in Ref. \cite{Cheng:yh}. According to their analysis,  the modified quantities due to nonlinear effects result in distinct corrections to the thermodynamic characteristics and phase structure of the EEH-AdS BHs in contrast to the RN-AdS BHs.
	
The structure of the paper is organized as follows: In Section \ref{action}, we briefly introduce the EEH-AdS/dS action and the corresponding field equations, and then we express the electrically charged EEH-AdS/dS BH solution. Section \ref{effectivemetric} is dedicated to obtaining the effective metric for the electrically charged EEH-AdS/dS BH. In Section \ref{optic},  we study the optical properties of the mentioned BHs and investigate the motion of photons in this spacetime by deriving the geodesic equations of the effective metric. In Subsection \ref{lightbending}, we give a full analysis on the trajectories of photon on
the equatorial plane. In Subsection \ref{shadow}, we calculate the radius of the BH shadow  and investigate the ratio
of shadow radius and photon sphere to find an acceptable optical behavior. In Subsection \ref{EHT}, we provide constraints on the parameters of the model using the recent observations from EHT on M87$^{*}$.  A study on the energy emission rate around the BH is presented in Subsection \ref{rate}. The gravitational deflection angle in the weak-field range is investigated in Subsection \ref{def}. We eventually summarize our results and conclude in Section \ref{conclusion}.
	\section{ ACTION AND FIELD EQUATIONS OF EULER-HEISENBERG THEORY}
	\label{action}
In this section, we give an overview of the Euler-Heisenberg BH since it is the base of our present work. The action of the EH theory coupled with gravity is given by \cite{Heisenberg:1936nmg}
	\begin{equation}\label{a1}
		S=\frac{1}{4\pi}\int_{M^{4}}d^{4}x\sqrt{-g}\left[\frac{1}{4}(R-2\Lambda)-\mathfrak{L}(\mathbb{F},\mathbb{G})\right],
	\end{equation}
	where $g$ and $R$ denote the metric tensor determinant and the Ricci scalar, respectively.	$\mathfrak{L}(\mathbb{F},\mathbb{G})$ is  the NLED Lagrangian defined as
	\begin{equation}\label{a2}
		\mathfrak{L}(\mathbb{F},\mathbb{G})=-\mathbb{F}+\frac{a}{2}\mathbb{F}^{2}+\frac{7a}{8}\mathbb{G}^{2},
	\end{equation}
	in which $\mathbb{F}=\frac{1}{4}\mathbb{F}_{\mu \nu} \mathbb{F}^{\mu \nu}$ and $\mathbb{G}=\frac{1}{4}\mathbb{F}_{\mu \nu}
{^*\mathbb{F}^{\mu \nu}}$ with $\mathbb{F}_{\mu\nu}$ denoting the electromagnetic field strength
tensor and ${^*\mathbb{F}^{\mu \nu}}=\epsilon _{\mu\nu\sigma\rho} \mathbb{F}^{\sigma\rho}
/(2\sqrt{-g})$ is its dual. $\epsilon_{\mu\nu\sigma\rho}$ is a completely antisymmetric  tensor that satisfies
$\epsilon_{\mu\nu\sigma\rho}\epsilon^{\mu\nu\sigma\rho}=-4!$. Moreover, $a=8 \alpha^2/45 m^4$ symbolizes the EH parameter which is employed to estimate the strength of the NLED correction;  $\alpha$ and $m$ are, respectively, the fine structure constant and electron mass. The case $ a=0 $ corresponds to Maxwell's theory of electrodynamics $\mathfrak{L}(\mathbb{F})=-\mathbb{F}$.

In NLED theory, there are two possible frameworks: $\mathbb{F}$ framework and  $\mathbb{P}$ framework. $\mathbb{F}$ framework is the classical frame wherein we
can use the $\mathbb{F}$ in terms of electromagnetic field tensor $\mathbb{F}^{\mu \nu}$. In $\mathbb{P}$ framework,  the tensor $\mathbb{P}_{\mu\nu}$ is used as the main field, defined by
	\begin{equation}\label{a3}
		\mathbb{P}_{\mu \nu}=-(\mathfrak{L}_{\mathbb{F}}\mathbb{F}_{\mu\nu}+{^*\mathbb{F}}_{\mu \nu}\mathfrak{L}_{\mathbb{G}}),
	\end{equation}
	where $ \mathfrak{L}_{\mathbb{F}} $ is the derivative of $ \mathfrak{L} $ w.r.t. $ \mathbb{F} $ and $ \mathfrak{L}_{\mathbb{G}} $ is the derivative of $ \mathfrak{L} $ w.r.t. $ \mathbb{G} $. In  the EH theory, $\mathbb{P}_{\mu \nu}$ takes the following
form,
	\begin{equation}\label{a4}
		\mathbb{P}_{\mu \nu}=(1-a\mathbb{F})\mathbb{F}_{\mu\nu}-{^*\mathbb{F}}_{\mu\nu}\frac{7a}{4}\mathbb{G}.
	\end{equation}
	
The two invariants, $\mathbb{P}$ and $\mathbb{O}$, associated with the $\mathbb{P}$ frame are given by	
	\begin{eqnarray}
	\mathbb{P}&=&-\frac{1}{4}\mathbb{P}_{\mu\nu}\mathbb{P}^{\mu\nu}, \\ \nonumber
	\mathbb{O}&=&-\frac{1}{4}\mathbb{P}_{\mu\nu}{^*\mathbb{P}}^{\mu\nu}, \\ \nonumber
	\end{eqnarray}
	where ${^*\mathbb{P}}_{\mu \nu}=\frac{1}{2\sqrt{-g}}\epsilon_{\mu\nu\sigma\rho}\mathbb{P}^{\sigma \rho}$. The Hamiltonian $\mathcal{H}$ can be defined by the Legendre transformation of $\mathfrak{L}$ as
	\begin{equation}\label{a7}
		\mathcal{H}(\mathbb{P},\mathbb{O})=\mathbb{P}-\frac{a}{2}\mathbb{P}^{2}-\frac{7a}{8}\mathbb{O}^{2}.
	\end{equation}

Using the aforementioned information, one can obtain the field equations as follows
	\begin{align}\label{a8}
		G_{\mu\nu}+\Lambda g_{\mu \nu}=8 \pi T_{\mu\nu} ,\,\ \nabla_{\mu}\mathbb{P}^{\mu \nu}=0.
	\end{align}
	
The energy-momentum tensor $T_{\mu \nu}$ in the $  \mathbb{P}$ frame takes the form	
	\begin{equation}\label{a9}
		T_{\mu \nu}=\frac{1}{4\pi}\left[(1-a\mathbb{P})\mathbb{P}^{\beta}_{\mu}\mathbb{P}_{\nu \beta}+g_{\mu \nu}\left(\mathbb{P}-\frac{3}{2}a\mathbb{P}^{2}-\frac{7a}{8}\mathbb{O}^{2}\right)\right].
	\end{equation}

The metric for EH-$ \Lambda $ BH presented in \cite{Heisenberg:1936nmg} is defined as	
	\begin{equation}\label{a10}
		ds^{2}=g_{\mu\nu}dx^{\mu}dx^{\nu}=-f(r)dt^{2}+f^{-1}(r)dr^{2}+r^{2}(d\theta^{2}+\sin^{2}\theta d\phi^{2}),
	\end{equation}
	in which
	\begin{equation}\label{a11}
		f(r)=1-\frac{2M}{r}+\frac{Q^{2}}{r^{2}}-\frac{\Lambda r^{2}}{3}-\frac{aQ^{4}}{20r^{6}},
	\end{equation}
	where $M$ and $Q$ are, respectively, the BH mass and its electric charge. $\Lambda$ is the cosmological constant that can be positive or negative. Note that with considering $a=0$, the BH reduces to the Riesner-Nordstrom (RN) solution with cosmological
constant (RN-$ \Lambda $). In addition, by setting $\Lambda=0$, the electrically charged EEH BH is recovered.

	\section{THE EFFECTIVE GEOMETRY FOR the electrically charged EEH-AdS/dS BLACK HOLE}
\label{effectivemetric}	
	
There is no interaction between electromagnetic waves and electrostatic fields in linear electrodynamics. However, in a nonlinear framework, a photon interacts with the field due to the field’s nonlinearity and moves along the null geodesics of an effective geometry instead of the Minkowski spacetime, implying that in a complex vacuum, light behaves like electromagnetic waves moving through a medium that alters their motion.  We will explore the method of \cite{Kruglov:ab,Novello:xy} to describe the effective geometry induced by EH NLED. Since the Lagrangian mentioned here is one-parameter, the equation of motion can be written as 
	\begin{equation}\label{a12}
		\partial_{\mu}(\mathfrak{L}_{\mathbb{F}}\mathbb{F}^{\mu \nu})=0.
	\end{equation}
	
The effective metric can be written in the following form

	\begin{equation}\label{a13}
		g_{ eff}^{\mu \nu}=\mathfrak{L}_{\mathbb{F}}g^{\mu \nu}-4\mathfrak{L}_{\mathbb{F}\mathbb{F}}\mathbb{F}^{\mu}_{\alpha}\mathbb{F}^{\alpha \nu},
	\end{equation}
	where $\mathfrak{L}_{\mathbb{F}}=d\mathfrak{L}/d\mathbb{F}$ and $\mathfrak{L}_{\mathbb{F}\mathbb{F}}=d^{2}\mathfrak{L}/d\mathbb{F}^{2}$. 
	Using equations \eqref{a12} and \eqref{a13}, and the metric element \eqref{a10}, the effective metric for the electrically charged EEH-AdS/dS BH is obtained as
	\begin{equation}\label{a14}
		ds_{ eff}^{2}=H(r)\left[-f(r)dt^{2}+f^{-1}(r)dr^{2}\right]+h(r) r^{2}( d\theta^{2}+\sin^{2}\theta d\phi^{2}),
	\end{equation}
where
	\begin{align}\label{a15}
		&H(r)=1+\frac{a(a Q^{3}-2Q r^{4})^{2}}{8 r^{12}}, \nonumber\\
		&h(r)=1+\frac{a(a Q^{3}-2Q r^{4})^{2}}{8 r^{12}}-\frac{a Q^{2}}{r^{4}}.
	\end{align}
	
	The resulting metric is the main equation we need to study the shadow behavior of EEH-AdS/dS BHs, which we will tackle in the next section.


	\section{Geodesics around EEH-AdS/dS Black hole}
	\label{optic}
In this section, we first determine the geodesic motion of photons in
the effective geometry characterized by Eq. (\ref{a14}). We then examine the optical features of the EEH-AdS/dS BH, such as the light ray trajectory, the radius of the photon sphere and shadow, energy emission rate, and gravitational lensing, and show how the mentioned quantities are affected by the parameters of the model. 
Additionally, we consider EEH-AdS/dS BHs as supermassive BHs and estimate the allowable ranges of the free parameters of the theory in light of the $ M87^{*} $  data.

	
	\subsection{Null geodesic and Light bending}
\label{lightbending}	
Now we are going to investigate the
geodesic structure of photons in a spherical symmetry spacetime. Since we work in a NLED framework, photons move along the null geodesic in the effective geometry. The Lagrangian of the EEH-AdS/dS BH spacetime is defined as
	\begin{equation}\label{a17}
		\mathcal{L}(x,\dot{x})=\frac{1}{2}g_{\mu \nu} \dot{x}^{\mu}\dot{x}^{\nu}=\frac{1}{2}[-H(r)f(r)\dot{t}^{2}+H(r)f^{-1}(r)\dot{r}^{2}+h(r)r^{2}(\dot{\theta}^{2}+\sin^{2}\theta \dot{\phi}^{2})],
	\end{equation}
where the dot notation represents differentiation with respect to an affine parameter,  denoted as $\lambda$. Since the BH under study is a spherically symmetric BH, the trajectory of a light outside a BH always lies on an equatorial plane $\theta=\frac{\pi}{2}$ and thereby the Lagrangian can be written as
	\begin{equation}\label{a18}
		\mathcal{L}=\frac{1}{2}[-H(r)f(r)\dot{t}^{2}+\frac{H(r)}{f(r)}\dot{r}^{2}]+h(r)\frac{1}{2}r^{2}\dot{\phi}^{2}.
	\end{equation}
	
From the Lagrangian, the canonical momentum can be obtained as	
\begin{eqnarray}
p_{t}&=&\frac{\partial \mathcal{L} }{\partial \dot{t}}= H(r) f(r) \dot{t}= E, \label{pt1} \\ 
p_{r}&=&\frac{\partial \mathcal{L}}{\partial \dot{r}}=\frac{-H(r)}{f(r)}\dot{r}, \label{pt2}\\
p_{\phi}&=&\frac{\partial \mathcal{L} }{\partial \dot{\phi}}= h(r) r^{2} \dot{\phi}~ sin^{2} \theta = L,
\label{pt3}
\end{eqnarray}
	in which $E$ and $L$ are the energy and orbital angular momentum
of the photon, respectively. Using the Lagrangian formalism,
the equations of motion for the
photons can be obtained in the following form
\begin{eqnarray}
\label{gm1}
&&\frac{{\rm d} t}{{\rm d} \lambda}=\frac{1}{b}H(r)^{-1}f(r)^{-1},\\
\label{gm2}
&&\frac{{\rm d} r}{{\rm d} \lambda}=\sqrt{\frac{1}{b^{2}H(r)^{2}} - \frac{1}{H(r)h(r)r^{2} }f(r)},\\
\label{gm3}
&&\frac{{\rm d} \phi}{{\rm d} \lambda}=\frac{1}{h(r) r^{2}},
\end{eqnarray}
where impact parameter $b = |L|/E$ is proportional to the closest distance of the light ray to the BH. From the Lagrangian (\ref{a18}) and Eqs. \eqref{pt1}, \eqref{pt2} and \eqref{pt3}, we can
write the following expression
\begin{equation}
\label{1-15}
H(r) f(r)\dot{t}^{2}-{H(r)}{f(r)^{-1}}\Big(\frac{{\rm d} r}{{\rm d} \phi}\Big)^{2}\dot{\phi}^{2} - r^{2} h(r) \dot{\phi}^{2} = 0.
\end{equation}

Utilizing Eq. (\ref{1-15}), one can calculate the effective potential $V_{\rm e}$ as 
\begin{equation}
\label{1-16}
\Big(\frac{{\rm d} r}{{\rm d} \phi}\Big)^{2} = V_{\rm e} = r^{4}\Bigg(\frac{h(r)^{2}}{b^{2}H(r)^{2}}-\frac{f(r)h(r)}{H(r)r^{2}}\Bigg).
\end{equation}

The analysis and investigation of the photon path near a BH depend on the initial conditions of the photon's trajectory, specifically the critical impact parameter. Due to the extremely high gravity around the BH, the photon can follow a circular orbit at a specific radius, known as the photon sphere, which can be calculated based on the properties of the BH and the effective potential. If the photon’s trajectory deviates slightly from the photon sphere, it will either fall into the BH or escape from it. With the usage of the effective potential (\ref{1-16}) and the condition $ d V_{\rm e}/dr=V_{\rm e}=0$,
one can determine the radius of the photon sphere ($r_{ph}$) and the critical impact parameter as
\begin{eqnarray}
\label{1-17}
&&\frac{1}{b^{2}}=\frac{E^{2}}{L^{2}}=\frac{f(r)H(r)}{r^{2}h(r)},\\
\label{1-18}
&&r f(r) H(r) h'(r) + 2 f(r) H(r) h(r)- r f(r) h(r) H'(r) - r H(r) h(r) f'(r)=0.
\end{eqnarray}

Then, by applying $r_{ph}$ obtained from equation \eqref{1-18} to the equation \eqref{1-17}, we obtain the critical impact parameter as follows
\begin{equation}\label{a26}
H(r_{ph})f(r_{ph})b_{c}^{2}=h(r_{ph})r_{ph}^{2}.
\end{equation}

By introducing a parameter $u \equiv {1}/{r}$, the orbit equation is obtained as
\begin{equation}
\left(\frac{{\rm d} u}{{\rm d} \phi} \right)^{2} =G(u),
\label{Gu}
\end{equation}
where
\begin{equation}
G(u)=\frac{h(u)^{2}}{b^{2}H(u)^{2}}-\frac{u^{2}f(u)h(u)}{H(u)}.
\label{Gu1}
\end{equation}

According to the impact parameter $b$, the photon trajectory near the BH can be classified into the following three situations:\\
(i) For $b>b_{c}$, the light beam approaches one nearest point of the BH and then returns to infinity. In this case, the turning point corresponds to the smallest positive real root of $ G(u)=0 $, denoted by $ u_{m} $. Using Eq. (\ref{Gu}),   the total change of azimuthal angle $ \phi $ reads
\begin{equation}
\phi=2\int_0^{u_m}\frac{du}{\sqrt{G(u)}},\quad b>b_c~.
\end{equation}

(ii) For $b<b_{c}$, the light ray falls into
the BH in all cases. Therefore, the total change of azimuthal angle $\phi$ is obtained by
\begin{equation}
\phi=\int_0^{u_0}\frac{du}{\sqrt{G(u)}},\quad b<b_c~,
\end{equation}
where $u_0=1/r_0$.

(iii) For  $b=b_{c}$, the photon enters an unstable circular orbit, called the photon sphere.\\
In Ref. \cite{Gralla:100}, the authors classified trajectories into three categories, direct, lensed, and photon rings to analyze the observational appearance of emission originating near a BH\footnote{In simple toy models with spherical accretion or isotropic emission, photon trajectories typically fall into three categories: direct, lensed, and photon ring rays. However, in thin-disk accretion models with equatorial emission extending to the horizon, an additional class of trajectories arises, those that do not cross the equatorial plane even once. These geodesics do not intersect the emitting region and thus appear dark in the image, giving rise to the so-called "inner shadow" feature, as discussed in \cite{Chael:918}.}. We provide a brief overview here. The total number of orbits can be defined as $n=\frac{\phi}{2\pi}$, which is a function of impact parameter $ b $, and satisfying \cite{Gralla:24018,Feng:5103} 
\begin{equation}
n(b)=\frac{2m-1}{4},\quad m=1,2,3,...~.
\label{nb}
\end{equation}

\begin{table}[ht]
	\caption{Regions of direct rays, lensing rings, and photon rings with $ M=1 $, $ a=0.1 $ and different values of $ Q $ and $ \Lambda $.}
	\label{table1}\centering
	\begin{tabular}{|c|c|c|c|}
		\hline\hline
		Parameters & $Q=0.2$, $ \Lambda =-0.01 $ & $Q=0.8$, $ \Lambda =-0.01$ & $Q=0.2$, $ \Lambda =-0.05$ \\[0.5ex] \hline
		Direct rays & $b<4.77$ & $b<4.17$ & $b<4.16$ \\ 
		$\left(n < \frac{3}{4}\right)$ & $b>5.78$ & $b>5.36$& $b>4.81$ \\ \hline
		Lensing rings & \quad$4.77<b<4.93$ & \quad $4.17<b<4.38$ & \quad $4.16<b<4.28$ \\ 
		$\left(\frac{3}{4}<n<\frac{5}{4}\right)$ & \quad $4.97<b<5.78$ & \quad $%
		4.43<b<5.36$ & \quad $%
		4.31<b<4.81$ \\ \hline
		Photon ring $\left(n>\frac{5}{4}\right)$ & \quad $4.93<b<4.97$ & \quad $%
		4.38<b<4.43$& \quad $%
		4.28<b<4.31$ \\[1ex] \hline
	\end{tabular}%
\end{table}
For each given $ m $, one can obtain two solutions $b_m^\pm$ for Eq. (\ref{nb}) where  $b_m^-<b_c$ and $b_m^+>b_c$. Following this, all trajectories can be classified as follows
\begin{itemize}
  \item direct: $n<\frac{3}{4}$, with $b\in(b_1^-,b_2^-)\cup(b_2^+,\infty)$. In this case, the photon trajectories intersect the equatorial plane only once.
  \item lensed: $\frac{3}{4} < n < \frac{5}{4}$, with $b\in(b_2^-,b_3^-)\cup(b_3^+,b_2^+)$. The photon trajectories intersect the equatorial plane at least twice.
  \item photon ring: $n>\frac{5}{4}$, with $ b\in(b_3^-,b_3^+) $. The photon trajectories intersect the equatorial plane at least three times.
\end{itemize}

In Table \ref{table1}, we illustrate the change in the range of $b$ values for direct emission, lensed ring emission, and photon ring emission of the EEH-AdS/dS BHs for different values of $Q$ and $ \Lambda $. It can be seen that the lensed rings and photon rings become thicker as the electric charge (cosmological constant) increases (decreases).

\begin{figure}[!htb]
	\centering
	\subfloat[ $Q=0.2 $, $ \Lambda =-0.01 $]{
		\includegraphics[width=0.31\textwidth]{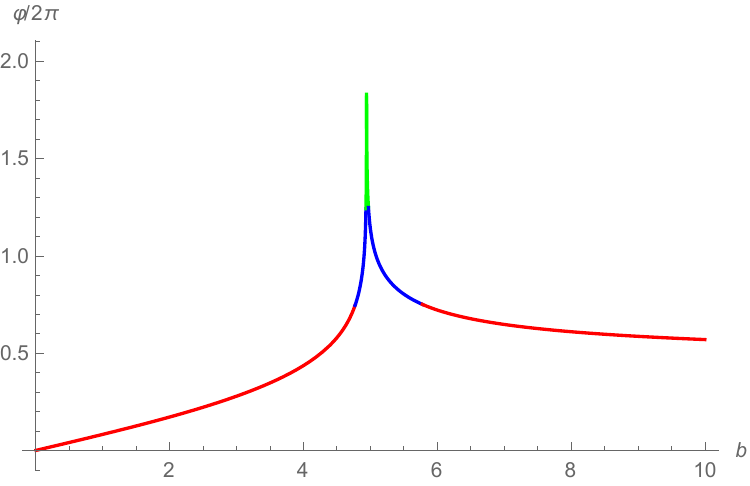}}
	\subfloat[ $Q=0.8 $, $ \Lambda =-0.01 $]{
		\includegraphics[width=0.31\textwidth]{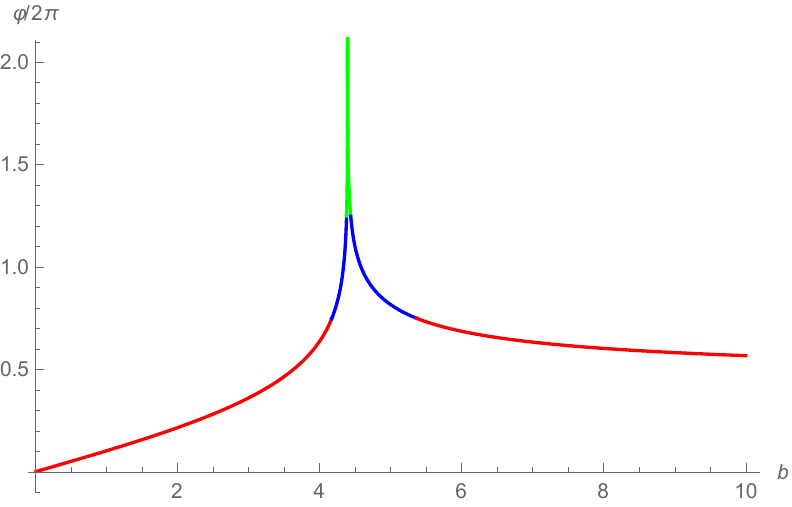}}
	\subfloat[ $Q=0.2 $, $ \Lambda =-0.05 $]{
		\includegraphics[width=0.31\textwidth]{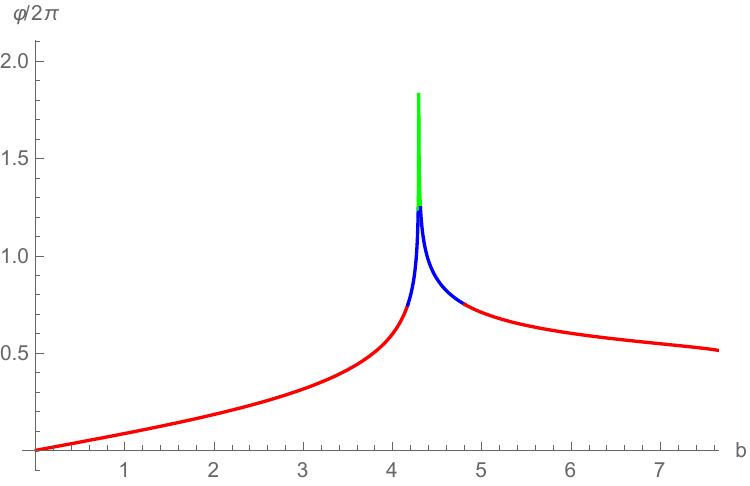}}        
	\newline
	\subfloat[ $Q=0.2 $, $ \Lambda =-0.01 $]{
		\includegraphics[width=0.31\textwidth]{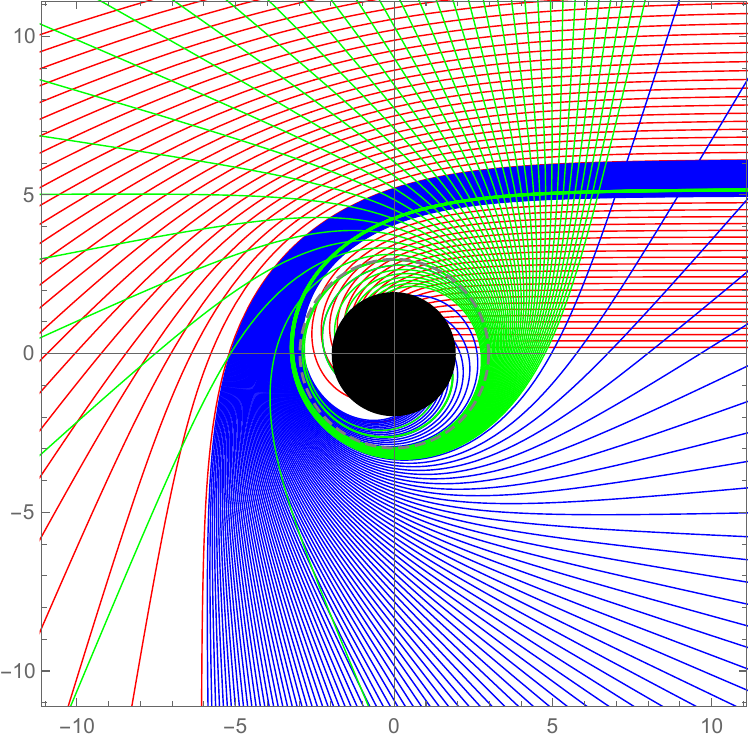}}
	\subfloat[ $Q=0.8 $, $ \Lambda =-0.01 $]{
		\includegraphics[width=0.31\textwidth]{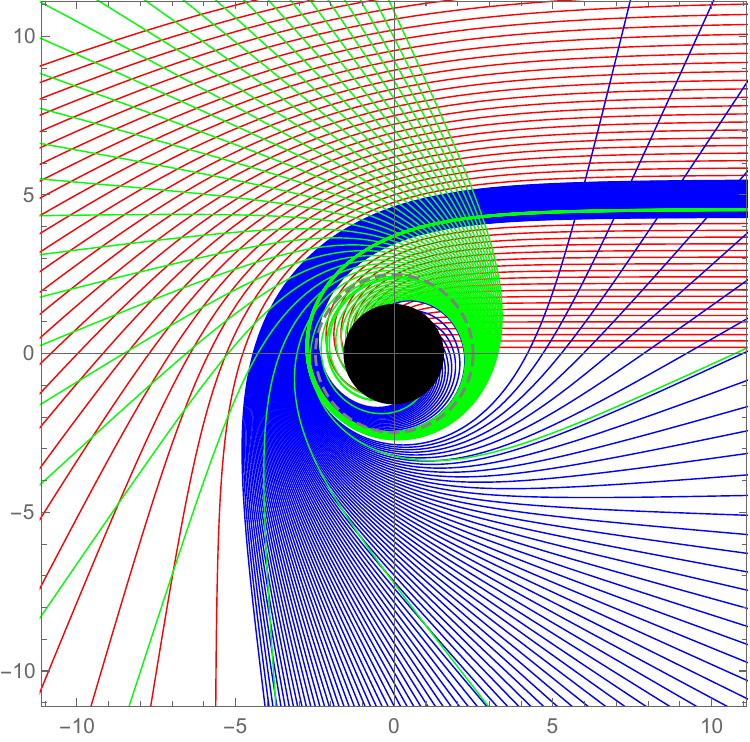}}
	\subfloat[ $Q=0.2 $, $ \Lambda =-0.05 $]{
		\includegraphics[width=0.31\textwidth]{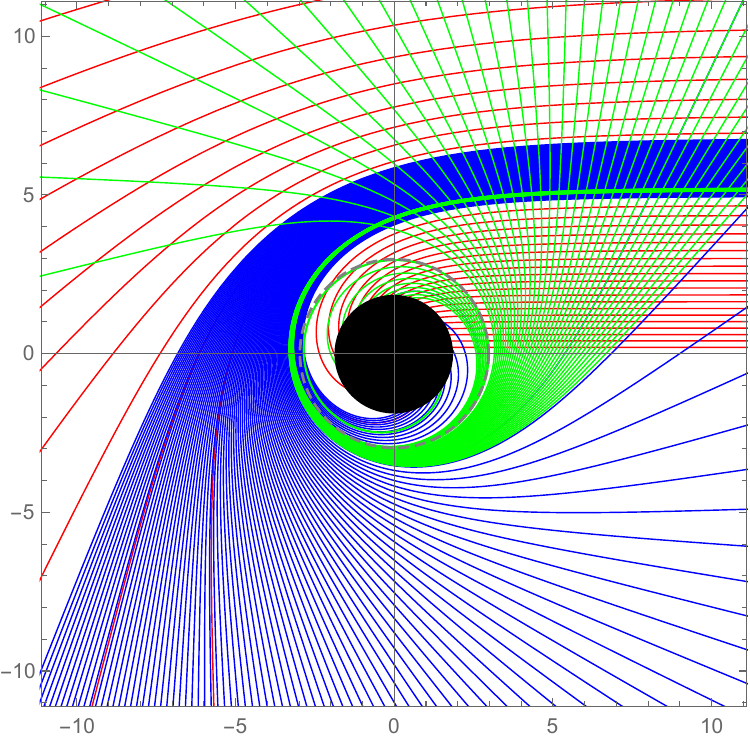}}        
	\newline
	\caption{The number of photon orbits $ n $ (top row) and trajectories of photons (bottom row) as a function of the impact
		parameter $ b $ for the EEH-AdS BH with $ a=0.1 $. The red lines, blue lines, and green lines correspond to $ (n<3/4) $, $ (3/4<n<5/4) $, and $ (n>5/4) $, respectively. 
The orbits are plotted in the equatorial plane ($\theta =\pi/2$),  represented as the horizontal plane, perpendicular to the vertical (polar) axis. 		
		The
		black disk and the dashed curves denote the event horizon and photon sphere.}
	\label{Figphoton}
\end{figure}
Fig. \ref{Figphoton} displays the number of photon
orbits $ n $ (top row) and trajectories of photons (bottom row) as a function of the
impact parameter $ b $. Red represents direct emission rays, blue denotes lensing rays, and green denotes photon ring rays. The photon orbit and the BH event horizon are shown by the dashed gray circle and black disk, respectively. From the top panels of Fig.\ref{Figphoton}, it is clear that when the impact parameter approaches a critical value $ b\pm b_{c} $, the photon trajectory exhibits a narrow peak in the $ (b,\phi) $ plane. Afterward, as $ b $ grows, the photon trajectories are always direct rays in any scenario. Comparing the middle and right panels of Fig. \ref{Figphoton} with the left panel of Fig. \ref{Figphoton}, one can find that both the electric charge and cosmological constant have a significant influence on the classification of light trajectories. From Table \ref{table1} and Fig. \ref{Figphoton}, one notices that increasing (decreasing) the electric charge (cosmological constant) results in broader ranges of photon and lensed ring emissions, depicted by the green and blue curves, respectively.

\subsection{Shadow of the EEH-AdS/dS BH}
	\label{shadow}
	BH shadow is the optical appearance cast by a BH in the sky. The shadow of a BH is a two-dimensional dark zone on the background of bright sources as seen by a distant observer. The boundary of the BH shadow delineates the visible image of the photon area by distinguishing between capture and scattering orbits. The photon region is effectively the edge of the spacetime region, which in the case of spherically symmetric spacetime corresponds to the photon sphere.	Inserting the photon sphere radius resulting from Eq. (\ref{1-18}) into $ V_{e}\left( r_{ph}\right)=0$, the shadow radius of the EEH-AdS/dS BH can be obtained as 
\begin{equation}\label{a28}
b_{c}=r_{sh}=r_{ph}\sqrt{\frac{h(r_{ph})}{H(r_{ph})f(r_{ph})}}.
\end{equation}

\begin{figure}[!htb]
\centering
\subfloat[]{
\includegraphics[width=0.31\textwidth]{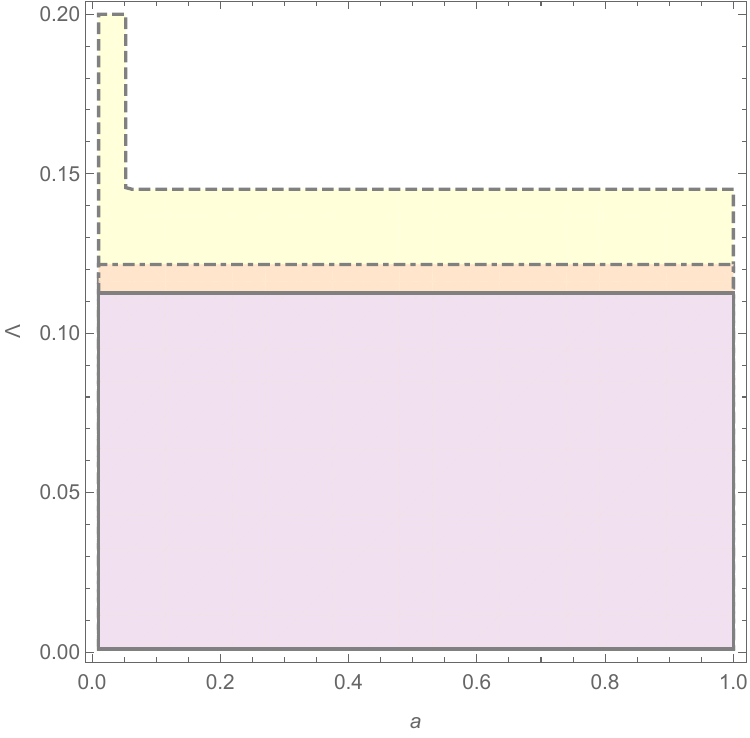}}
\subfloat[]{
\includegraphics[width=0.31\textwidth]{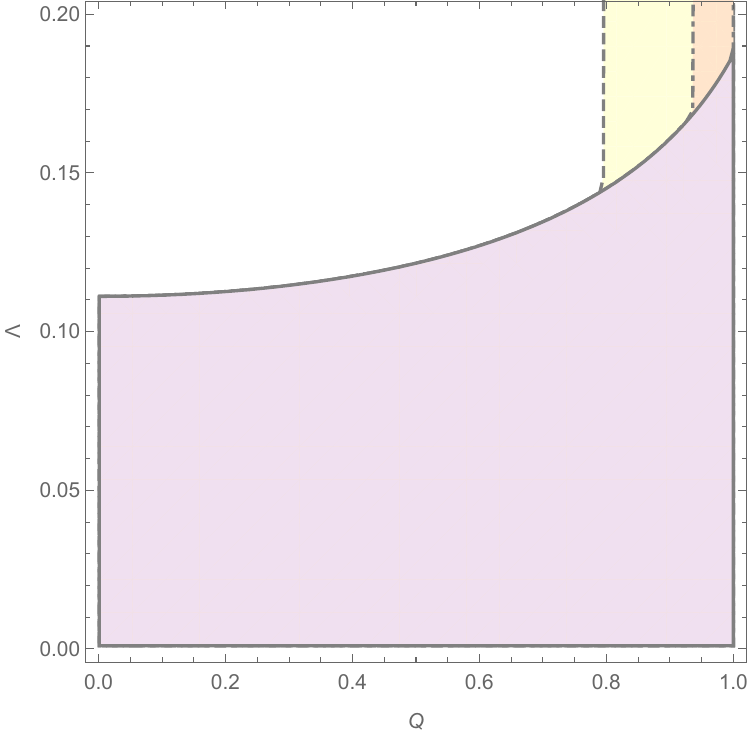}}        
\newline
\subfloat[]{
\includegraphics[width=0.31\textwidth]{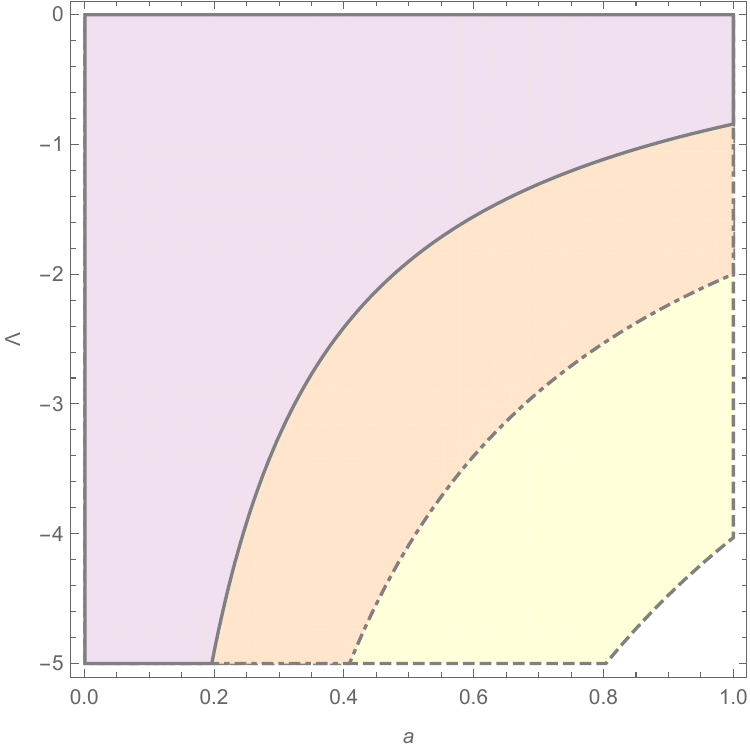}}
\subfloat[]{
\includegraphics[width=0.31\textwidth]{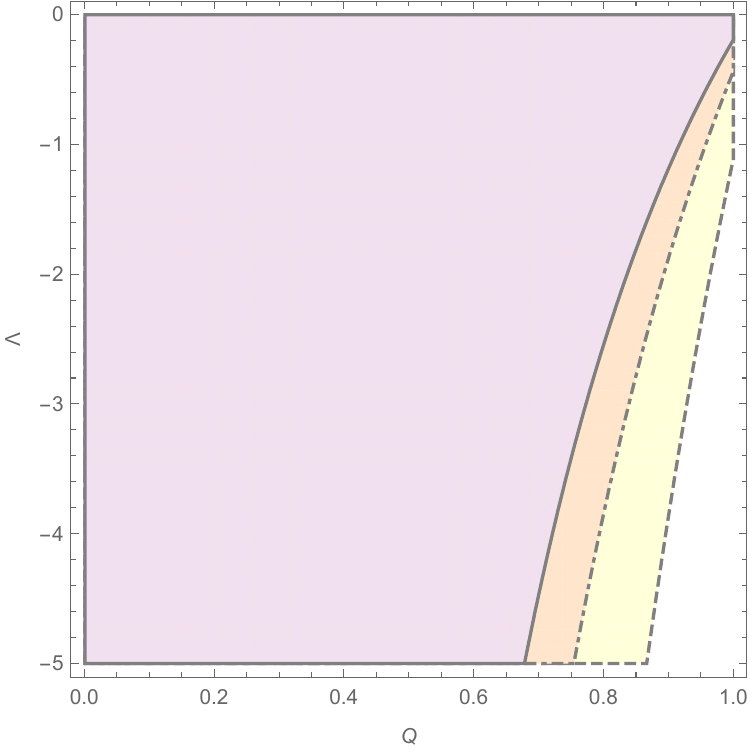}}        
\newline
\caption{\textbf{Top row:} The constraint $\frac{r_{ph}}{r_{eh}}>1$ (shaded areas) for the EEH-dS BH is displayed in (a): $(\Lambda,a)$ plane for $ Q=0.2$ (continuous curve), $Q=0.5$ (dashdotted curve) and $Q=0.8$ (dashed curve); (b): $(\Lambda,Q)$ plane for  $a=0.4$ (continuous curve), $a=0.2$ (dashdotted curve) and $a =0.05$ (dashed curve). \textbf{Bottom row:} The constraint $\frac{r_{ph}}{r_{eh}}>1$ (shaded areas) for the EEH-AdS BH is displayed in (a): $(\Lambda,a)$ plane for $ Q=0.9$ (continuous curve), $Q=0.8$ (dashdotted curve) and $Q=0.7$ (dashed curve); (b): $(\Lambda,Q)$ plane for  $a=0.9$ (continuous curve), $a=0.6$ (dashdotted curve) and $a =0.3$ (dashed curve). Here, the colorless regions represent the areas where $\frac{r_{ph}}{r_{eh}}<1$ which is physically forbidden.}
\label{Fig1}
\end{figure}
\begin{figure}[!htb]
\centering
\subfloat[]{
\includegraphics[width=0.31\textwidth]{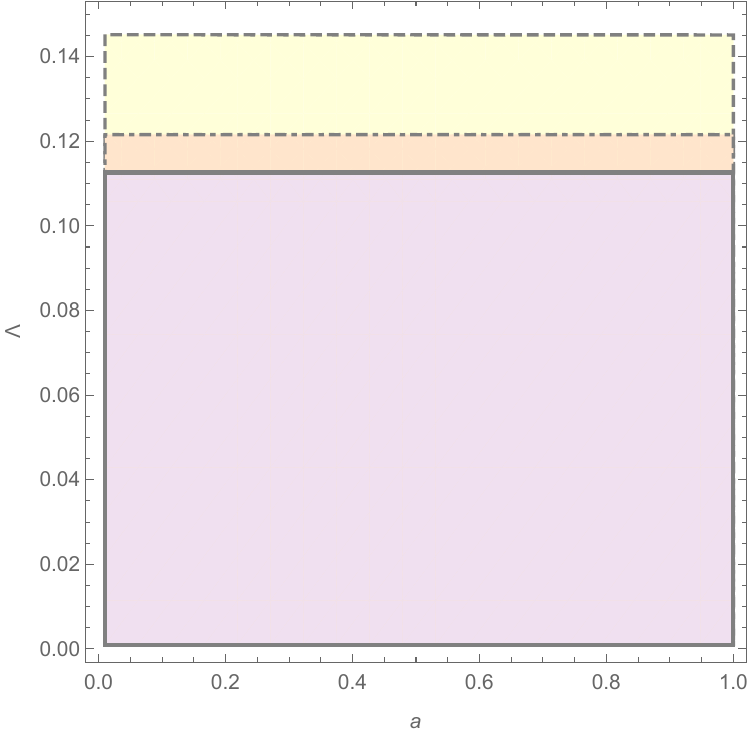}}
\subfloat[]{
\includegraphics[width=0.31\textwidth]{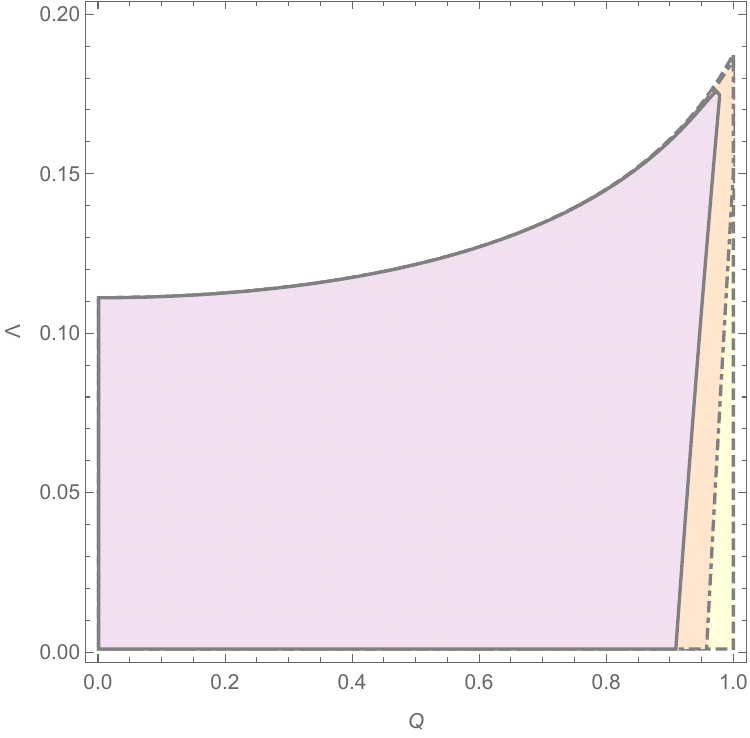}}        
\newline
\subfloat[]{
\includegraphics[width=0.31\textwidth]{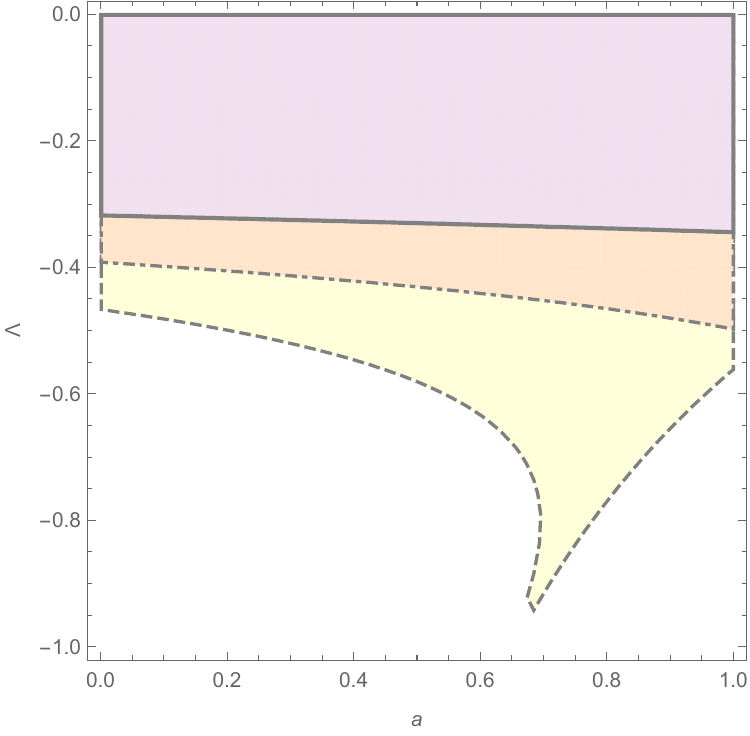}}
\subfloat[]{
\includegraphics[width=0.31\textwidth]{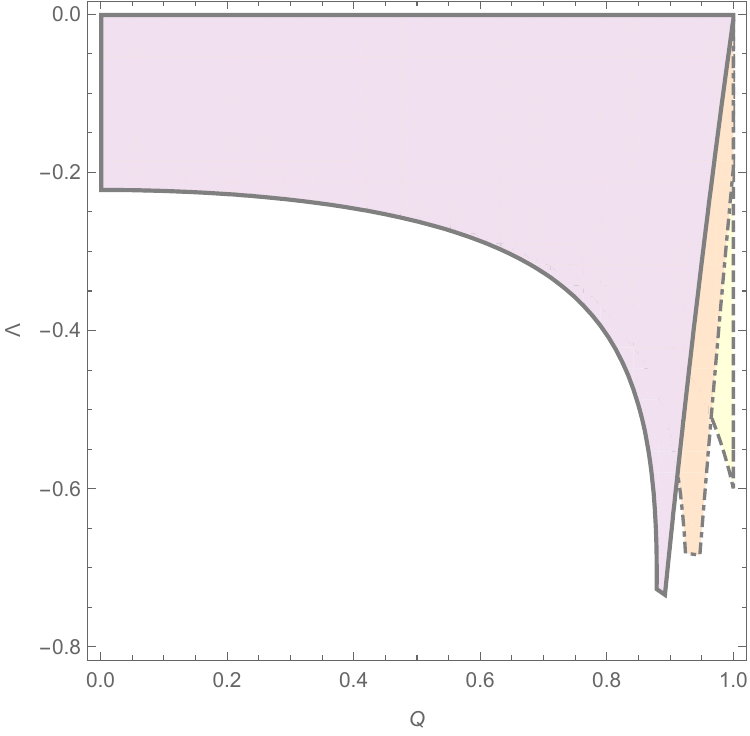}}        
\newline
\caption{\textbf{Top row:} The constraint $\frac{r_{sh}}{r_{ph}}>1$ (shaded areas) for the EEH-dS BH is displayed in (a): $(\Lambda,a)$ plane for $ Q=0.2$ (continuous curve), $Q=0.5$ (dashdotted curve) and $Q=0.8$ (dashed curve); (b): $(\Lambda,Q)$ plane for  $a=1.0$ (continuous curve), $a=0.5$ (dashdotted curve) and $a =0.01$ (dashed curve). \textbf{Bottom row:} The constraint $\frac{r_{sh}}{r_{eh}}>1$ (shaded areas) for the EEH-AdS BH is displayed in (a): $(\Lambda,a)$ plane for $ Q=0.75$ (continuous curve), $Q=0.85$ (dashdotted curve) and $Q=0.95$ (dashed curve); (b): $(\Lambda,Q)$ plane for  $a=1.0$ (continuous curve), $a=0.5$ (dashdotted curve) and $a =0.01$ (dashed curve). Note that in the unshaded (colorless) regions the condition $\frac{r_{sh}}{r_{ph}}<1$ is violated.}
\label{Fig2}
\end{figure}
In general,  $ b_{c} $ represents the apparent radius of the shadow cast by a static, spherically symmetric black hole as seen by a distant observer \cite{Perlick:947,Qiao:106,Chena:512}. To observe an acceptable optical behavior, we need to examine the condition $r_{eh} < r_{ph} < r_{sh}$, where $ r_{eh} $ is  the event horizon radius. Such investigation helps us find the allowed ranges of parameters to have an acceptable physical result. Fig. \ref{Fig1} displays the admissible parameter space which satisfies the constrain $\frac{r_{ph}}{r_{eh}}>1$. The colored area is where the mentioned constraint holds. The top row indicates the admissible regions of parameters for EEH-dS BHs. From Fig. \ref{Fig1} (a), it can be seen that the admissible regions increases as the electric charge increases. While the EH parameter $ a $ decreases the admissible regions (see Fig. \ref{Fig1} (b)). Regarding the EEH-AdS BH, both parameters $ Q $ and  $ a $ have decreasing contribution to the admissible parameter space (see bottom panels of Fig. \ref{Fig1}). 

To investigate the constrain $\frac{r_{sh}}{r_{ph}}>1$, we have plotted Fig. \ref{Fig2}. The top row displays the admissible regions of parameters for EEH-dS BHs. As can be seen, the electric charge increases the acceptable areas (see Fig. \ref{Fig2} (a)), while according to Fig. \ref{Fig2} (b), the parameter $a$ decreases it. For the case of  EEH-AdS BHs, the bottom panels of Fig. \ref{Fig2} confirm that enhancing both parameters leads to an increase in the acceptable area.

\begin{figure}[!htb]
\centering
\subfloat[ $ Q=0.8 $ and $ \Lambda=0.02$]{
\includegraphics[width=0.31\textwidth]{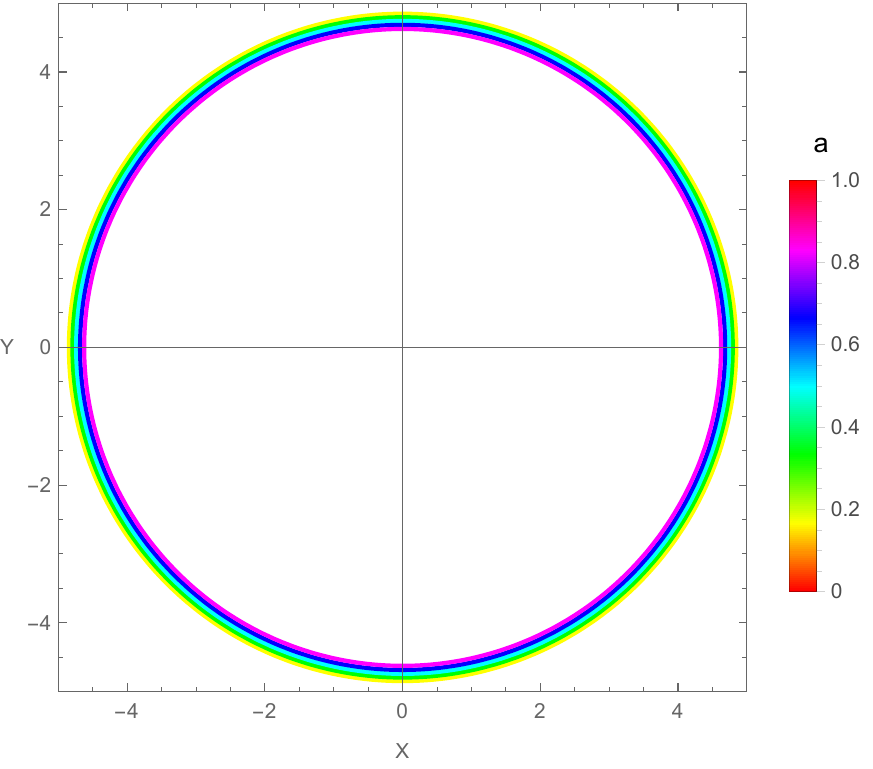}}
\subfloat[$ a=0.2 $ and $ \Lambda=0.02$]{
     \includegraphics[width=0.31\textwidth]{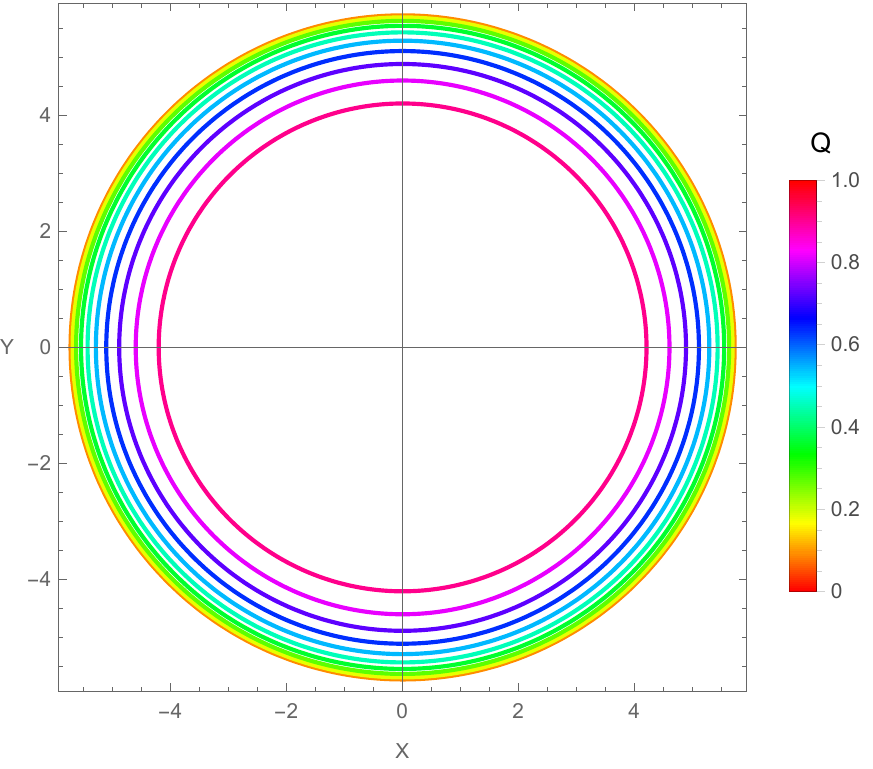}}       
\newline
\subfloat[$ a=0.2 $ and $ Q=0.5$]{
\includegraphics[width=0.31\textwidth]{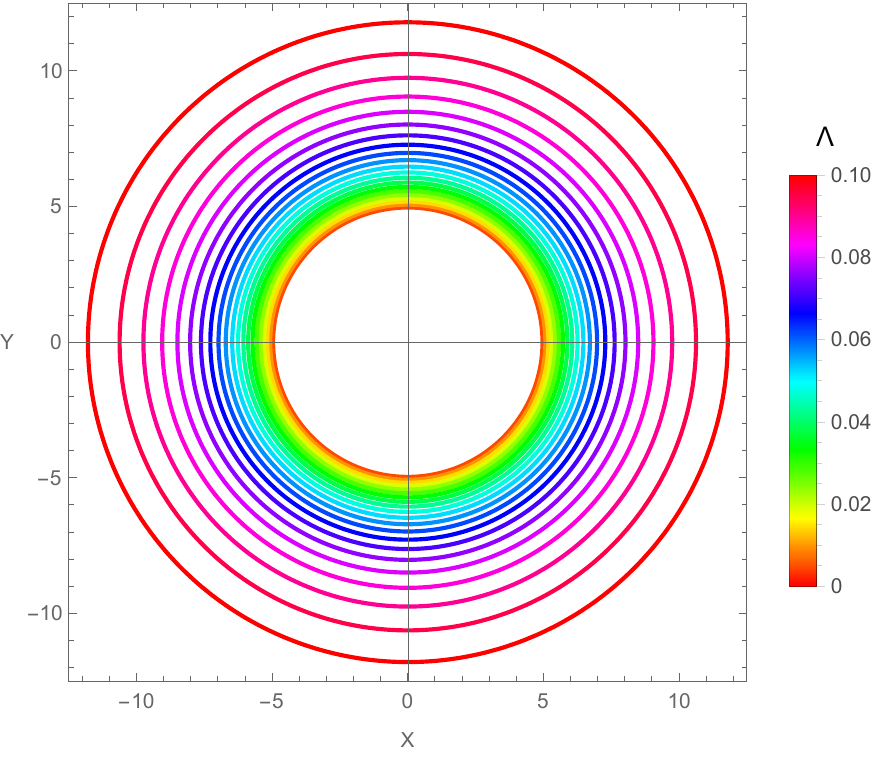}}
\subfloat[$ a=0.2 $ and $ Q=0.5$]{
     \includegraphics[width=0.31\textwidth]{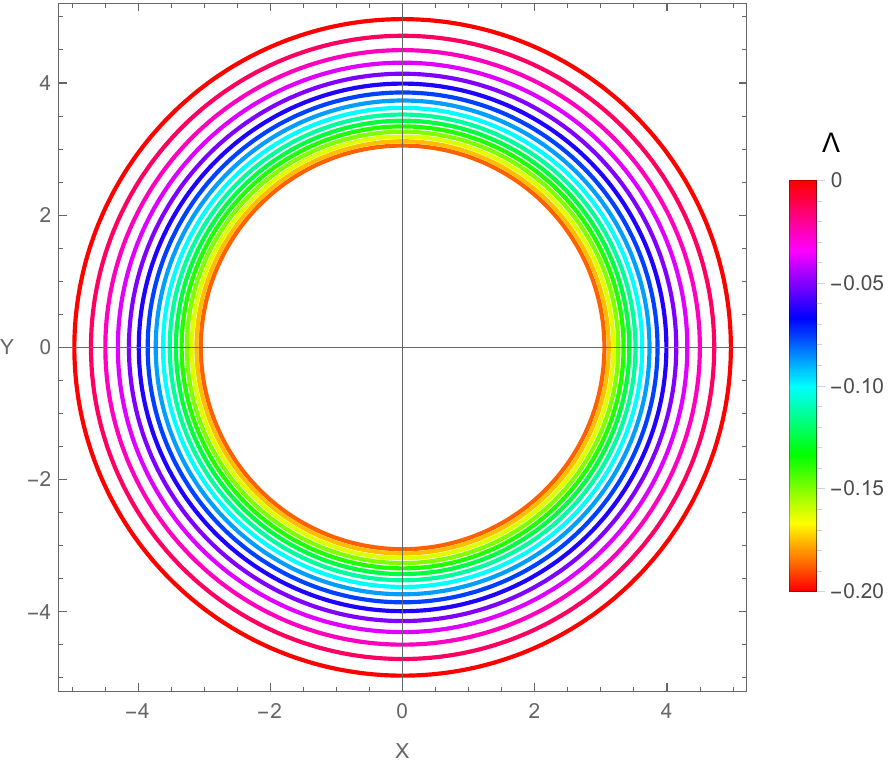}}        
\newline
\caption{The BH shadow in the Celestial plane $(X-Y)$ for different values of the parameters  $ a $, $ Q $ and $ \Lambda$.}
\label{Fig4}
\end{figure}

Fig. \ref{Fig4} shows how BH parameters affect the size of the BH shadow. To study the influence of the EH parameter, we plotted Fig. \ref{Fig4}(a) while keeping fixed values for the electric charge $Q$ and the cosmological constant $ \Lambda $ and found that the shadow size shrinks with increasing $ a $. This result is consistent with previous studies in the context of NLED, which indicate that the shadow size generally decreases with increasing NLED field strength. For example, Ref. \cite{W110} examined the shadows of Bardeen and Ghosh-Culeta BHs, showing a reduction in shadow size as the NLED field intensifies. Similarly, Ref. \cite{Okyay:009} analyzed spherically symmetric BHs within the double logarithmic NLED framework and observed a slight decrease in shadow size with increasing NLED parameters. The influence of NLED on rotating BHs was further studied in Ref. \cite{Ahmedov:2016}, where it was found that the shadows of rotating Hayward and Bardeen BHs shrink in the presence of NLED effects.  Additionally, Ref. \cite{Kumar:100} reported that for a rotating charged Hayward BH, a stronger NLED field not only reduces the shadow size but also increases its distortion.

The impact of the electric charge on the shadow size is illustrated in Fig. \ref{Fig4}(b) with fixed values of $ a $ and $ \Lambda $, illustrating that the parameter $Q$ has a reducing effect on $r_{sh}$. In Figs. \ref{Fig4}(c) and \ref{Fig4}(d), we set $a$ and $Q$ as fixed parameters and investigated the change in the shadow size with the variation of the cosmological constant. According to our findings, for EEH-dS BHs, the cosmological constant has an increasing contribution to the shadow size. Meanwhile, for EEH-AdS BHs, the radius of shadow shrinks with the growth of $\vert \Lambda \vert$.

	\subsection{Constraints from EHT Observation of $M87^{*}$}
	\label{EHT}
The latest observations from the EHT of the supermassive BHs $M87^{*}$ have created a new scientific opportunity for testing gravity theories through BH shadow observables and significantly advanced our understanding of gravitational physics. In literature, the BH shadows are modeled and constrained using BH candidates and observational data. In this subsection, we are going to model $M87^{*}$ as the EEH-AdS/dS BHs and impose the EHT-inferred bounds on the
 shadow diameter ($ d_{sh} $) and Schwarzschild shadow deviation $ \delta $ to find the constraints on the parameters. 		

EHT collaboration reported that, for supermassive BH $M87^{*}$, the angular diameter of the shadow $ \theta_{M87^{*}} $, the mass $ M_{M87^{*}} $, and the distance of the $M87^{*}$ from
the Earth $ D_{M87^{*}} $ are as follows \cite{EventHorizonTelescope:2019dse}
	\begin{eqnarray}
	\theta_{M87^{*}} &=& (42 \pm 3) ~\mu c~,\\ \nonumber
	M_{M87^{*}}&=&(6.5 \pm 0.9) \times 10^{9}M_{\odot}~,\\ \nonumber
	D_{M87^{*}}&=&16.8_{-0.7}^{+0.8} ~ Mpc~,
	\end{eqnarray}
	where $M_{\odot}$ is Sun mass. From these reported values, the shadow diameter in units of mass is bounded as
 \cite{EventHorizonTelescope:2019ths}
	\begin{equation}
		d_{M87^{*}}\equiv\frac{D \theta}{M} \approx 11.0 \pm 1.5~.
		\label{dM87}
	\end{equation}
	
As can be seen from Eq. (\ref{dM87}), $9.5  \lesssim d_{M87^{*}}  \lesssim 12.5$ within $ 1\sigma $ uncertainty, whereas within $ 2\sigma $ uncertainty $ 8 \lesssim d_{M87^{*}} \lesssim 14 $. 

\begin{figure}[!htb]
\centering
\subfloat[ $ \Lambda=0.035 $]{
\includegraphics[width=0.31\textwidth]{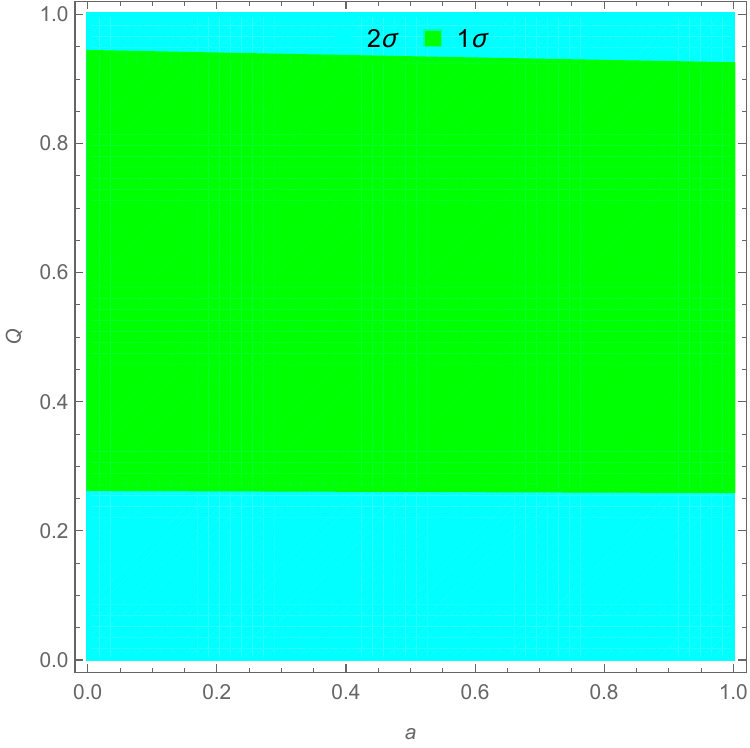}}
\subfloat[$ Q=0.5 $]{
     \includegraphics[width=0.31\textwidth]{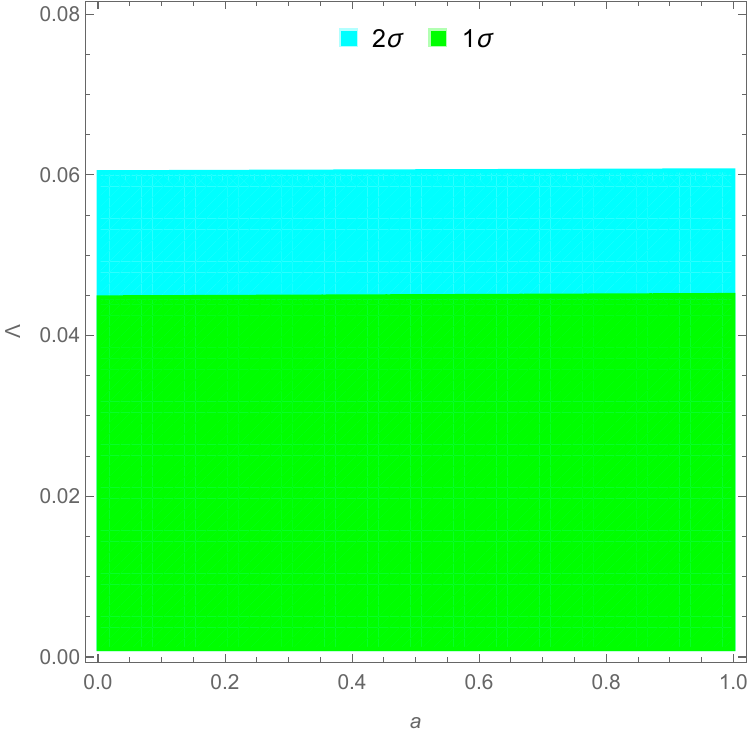}}
     \subfloat[$ a=0.2 $]{
     \includegraphics[width=0.31\textwidth]{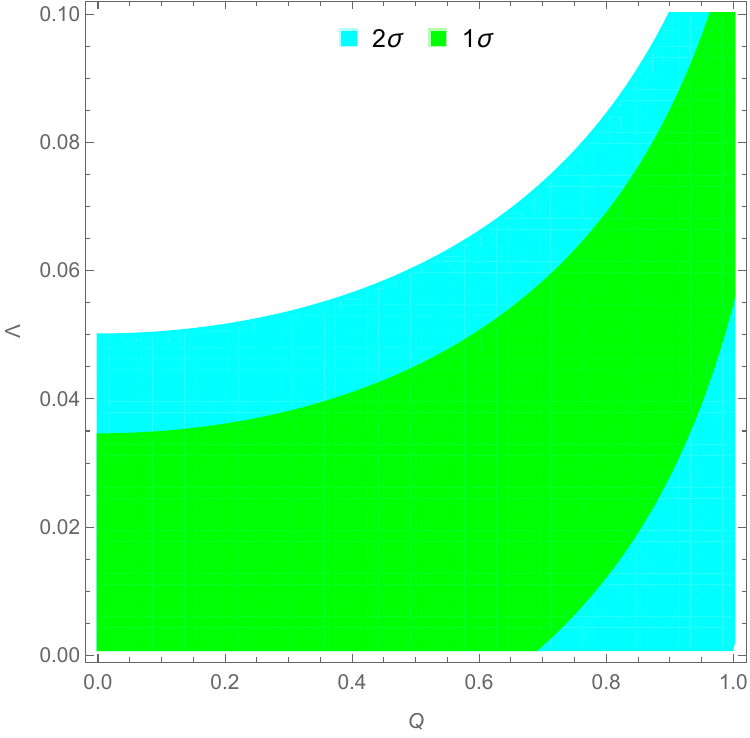}}               
\newline
\subfloat[$ \Lambda=-0.01 $]{
\includegraphics[width=0.31\textwidth]{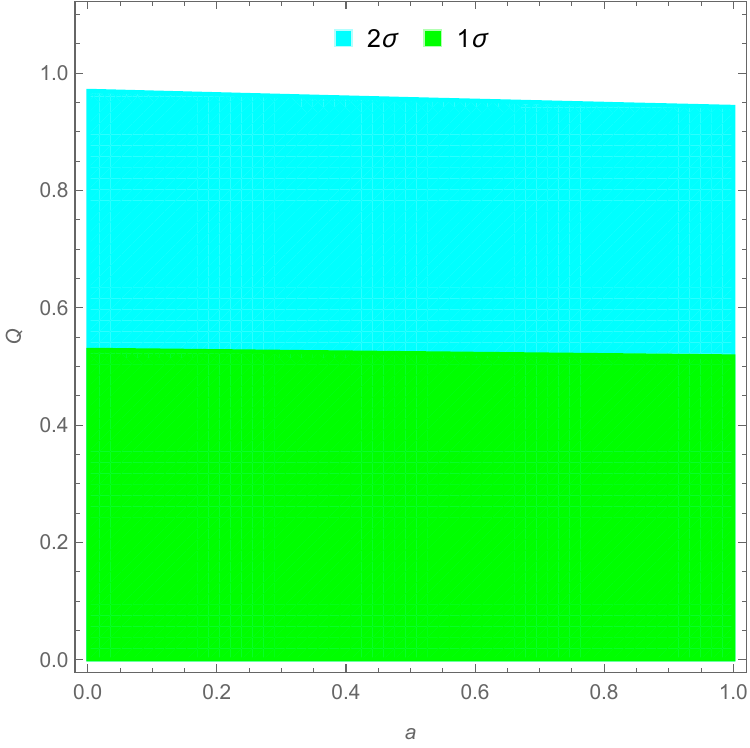}}
\subfloat[$ Q=0.5 $]{
     \includegraphics[width=0.31\textwidth]{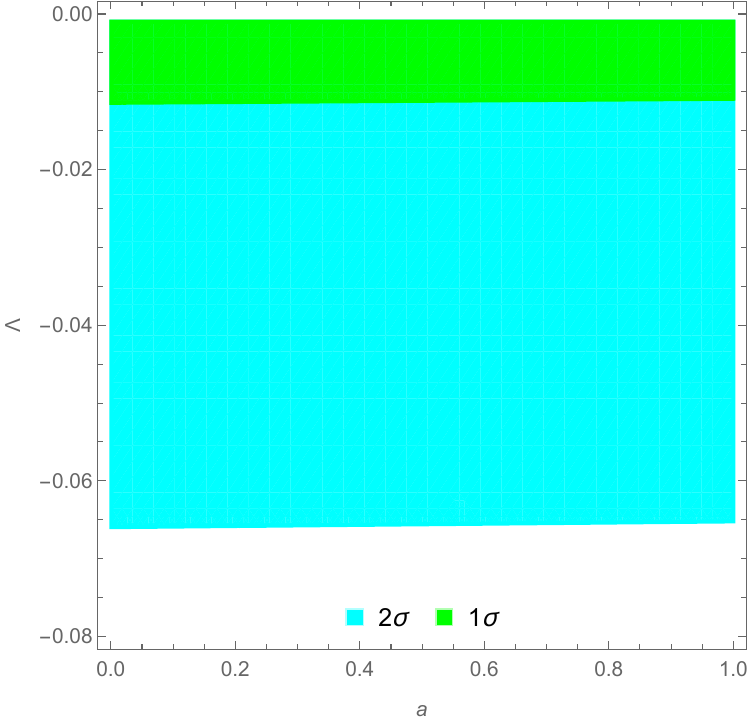}}
     \subfloat[$ a=0.2 $]{
     \includegraphics[width=0.31\textwidth]{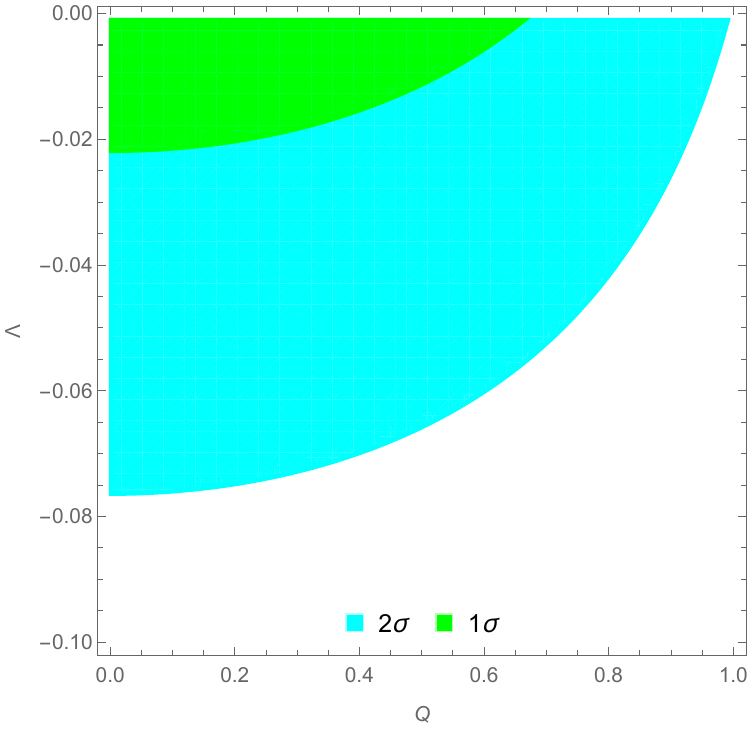}}        
\newline
\caption{Constraints on parameters of the EEH-dS BH (top row) and EEH-AdS BH (bottom row) with the EHT observations
of $ M87^{*} $.}
\label{FigEHT}
\end{figure}

Fig. \ref{FigEHT} shows the allowed regions of the model's parameters for the EEH-dS BH (top row) and EEH-AdS BH (bottom row) which satisfies the constraint (\ref{dM87}) within $ 1\sigma $ and $ 2\sigma $ uncertainty. The green shaded regions corresponds to the $ 1\sigma $ confidence region for $ d_{sh} $, whereas the
cyan shaded regions gives the $ 2\sigma $ confidence region. 
Figs. \ref{FigEHT}(a) and \ref{FigEHT}(d) show the constraints on parameter space of $(a,Q) $  for a fixed cosmological constant. In dS spacetime, the resulting shadow lies within $ 1\sigma $ uncertainty
for $0.25 <Q<0.95 $ and within $ 2\sigma $ uncertainty for $ Q<0.25 $ and $ Q>0.95 $ (see Fig. \ref{FigEHT}(a)). While in AdS spacetime, $ d_{sh} $ is consistent with
observational data within $ 1\sigma $-error for $ Q<0.53 $ and $ 2\sigma $-error for $0.53 <Q<0.97 $ (see Fig. \ref{FigEHT}(d)). In Figs. \ref{FigEHT}(b) and \ref{FigEHT}(e), we set $ Q=0.5 $ and determine the allowable range of the cosmological constant. According to Fig. \ref{FigEHT}(b), the shadow diameter of the EEH-dS BH is located in $ 1\sigma $ confidence region for $ \Lambda <0.045 $ and $ 2\sigma $ confidence region for $0.045 <\Lambda <0.06 $. As for EEH-AdS BHs, the range $ \Lambda > -0.012 $ can satisfy $ 1\sigma $ bound, whereas the range $ -0.066 < \Lambda < -0.012 $ satisfies $ 2\sigma $ bound (see Fig. \ref{FigEHT}(e)). Regarding the allowed range of the EH parameter, it can be seen from Fig. \ref{FigEHT} that the constraint \eqref{dM87} is satisfied for all values of this parameter.
Comparing the top and bottom panels of Fig. \ref{FigEHT} with each other, one can find that the EEH-dS BH can be considered as a suitable candidate for $M87^{*}$ supermassive BH.  

\begin{figure}[!htb]
\centering
\subfloat[ $ \Lambda=0.035 $]{
\includegraphics[width=0.31\textwidth]{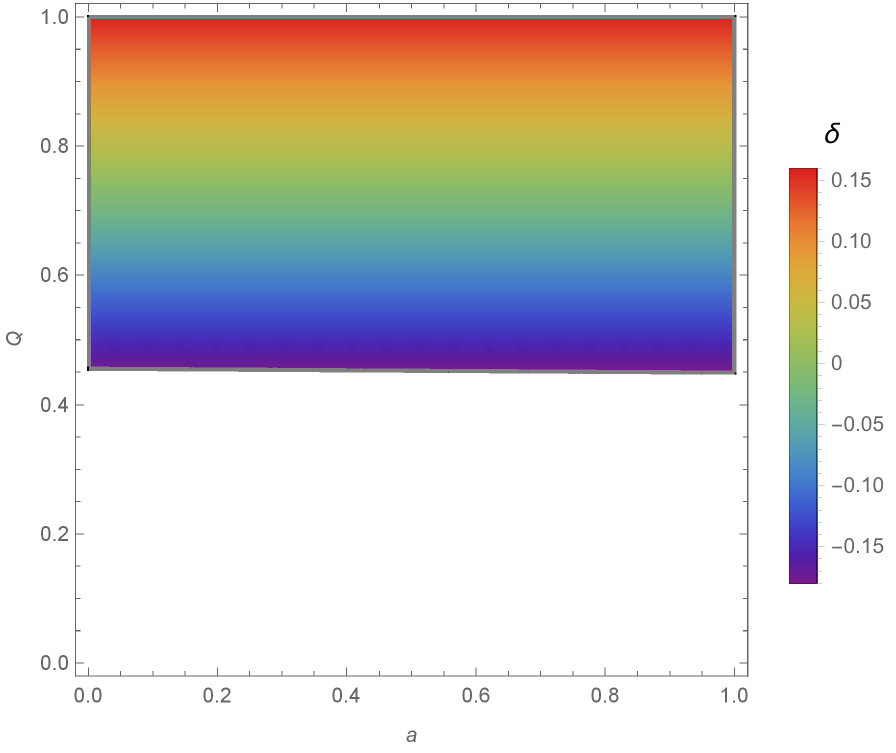}}
\subfloat[$ Q=0.5 $]{
     \includegraphics[width=0.31\textwidth]{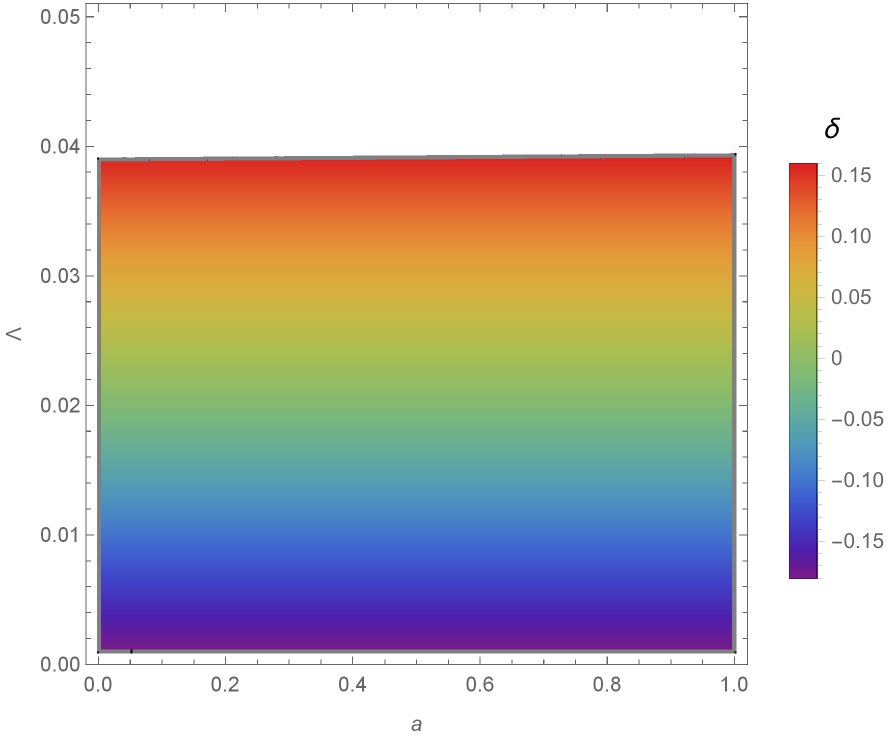}}
     \subfloat[$ a=0.2 $]{
     \includegraphics[width=0.31\textwidth]{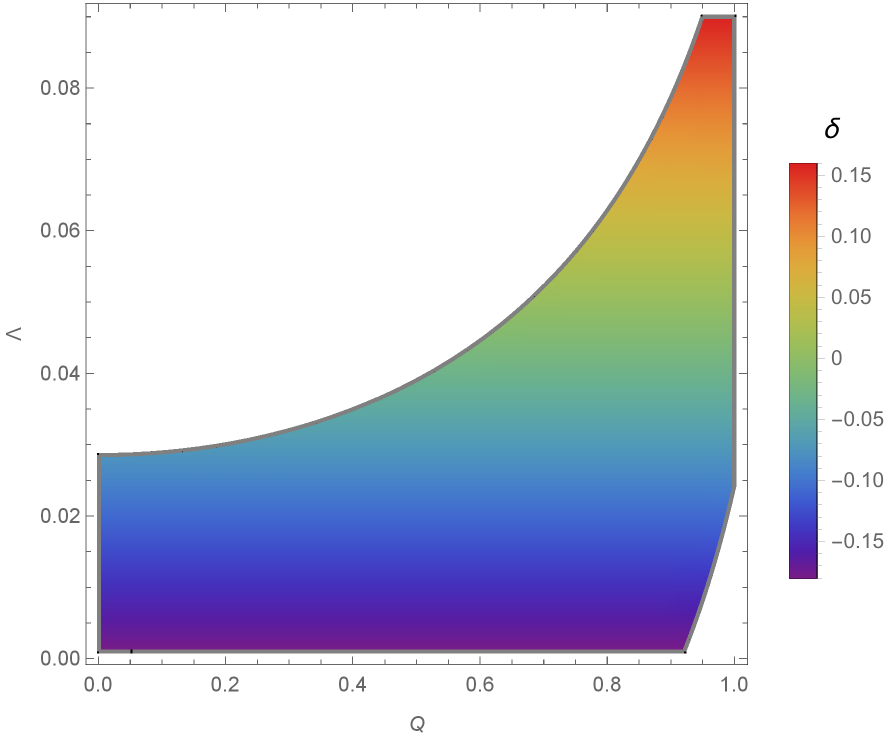}}               
\newline
\subfloat[$ \Lambda=-0.01 $]{
\includegraphics[width=0.31\textwidth]{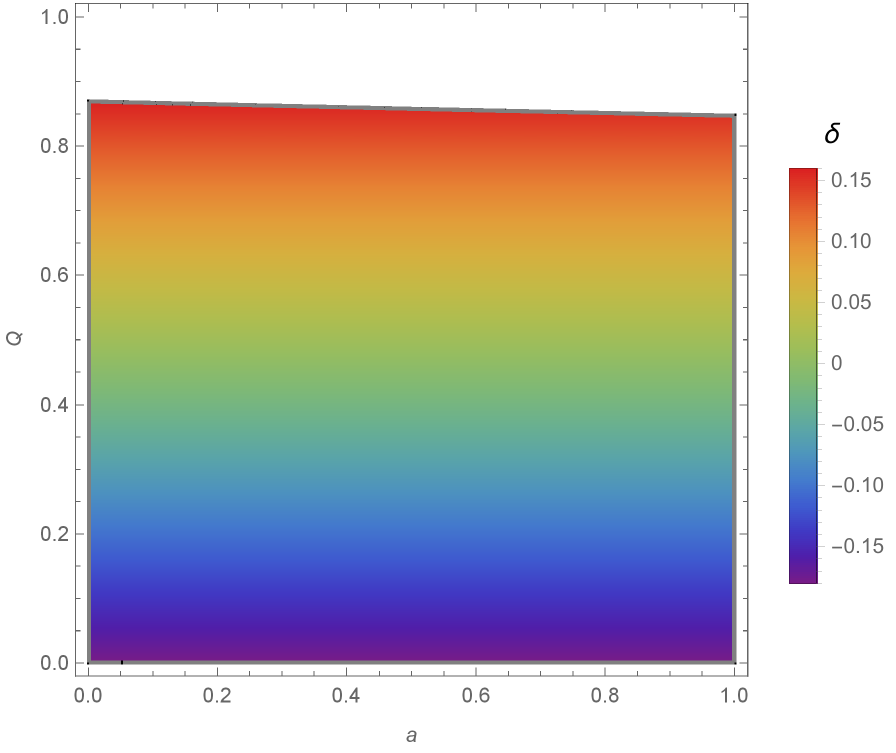}}
\subfloat[$ Q=0.5 $]{
     \includegraphics[width=0.31\textwidth]{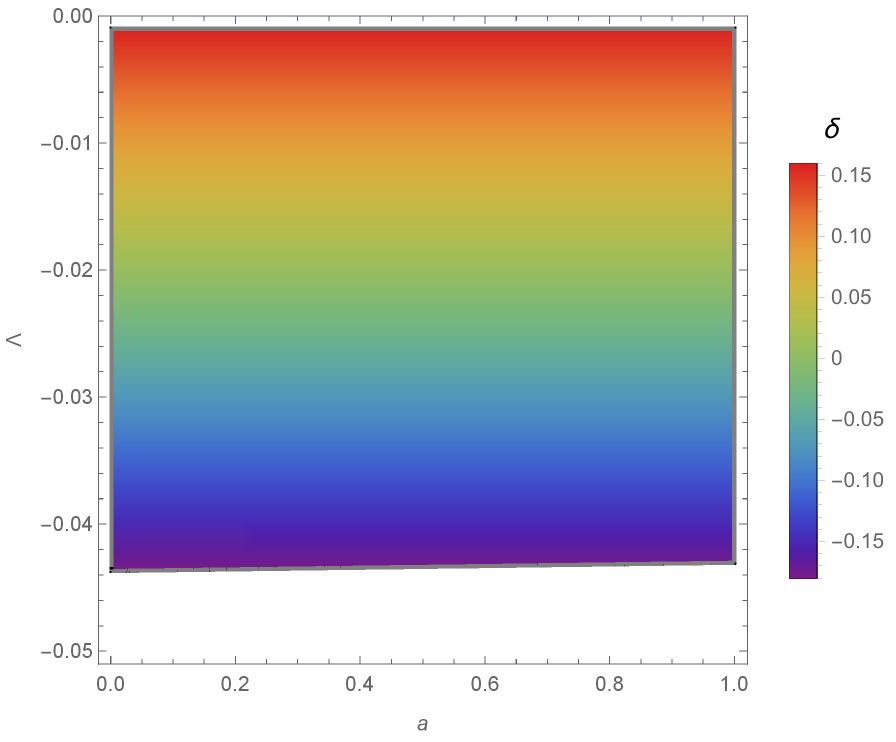}}
     \subfloat[$ a=0.2 $]{
     \includegraphics[width=0.31\textwidth]{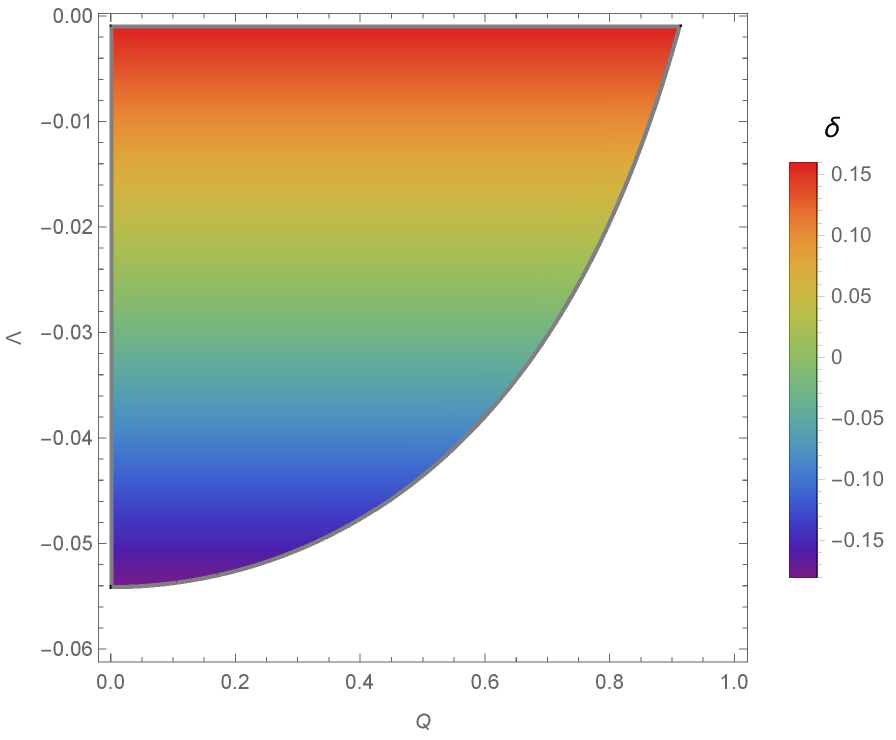}}        
\newline
\caption{\textbf{Top row:} The difference between shadow diameter of the EEH-dS BH and Schwarzschild BH as a function of $(Q,a)$, $ (\Lambda,a) $ and $  (\Lambda,Q) $. \textbf{Bottom row:} The difference between shadow diameter of the EEH-AdS BH and Schwarzschild BH as a function of $(Q,a)$, $ (\Lambda,a) $ and $  (\Lambda,Q) $. Note
that the colorless regions are forbidden for $  a$, $  Q$  and $ \Lambda $.}
\label{Figdelta}
\end{figure}

EHT collaboration estimated another important observable, known as the Schwarzschild shadow deviation ($ \delta $) which measures the difference between the model shadow diameter ($ d_{metric} $) and the Schwarzschild BH shadow diameter as \cite{EventHorizonTelescope:2019ths}
	\begin{equation}\label{d3}
		\delta=\frac{d_{metric}}{6\sqrt{3}}-1,
	\end{equation}
where $d_{metric} = 2r_{sh}$. Based on $M87^{*}$ observation, the  Schwarzschild shadow deviation is bounded as $-0.18< \delta < 0.16$. Obviously, if the size of the BH shadow is larger (smaller) than a Schwarzschild BH of the same mass, $\delta$ will be positive (negative). The shaded area in Fig. \ref{Figdelta} illustrates the admissible parameter space for the EEH-dS BH (top row) and the EEH-AdS BH (bottom row) for which the mentioned constraint for $\delta$ is satisfied. From Fig. \ref{Figdelta} (a), it can be seen that in dS spacetime and for the fixed cosmological constant, the constraint on $\delta$ is satisfied for the range $Q>0.45$. Furthermore, this figure shows that the size of the shadow becomes larger than the Schwarzschild BH shadow for very large values of electric charge.  The allowed range of the cosmological constant is depicted in Fig.  \ref{Figdelta} (b) for fixed $Q$ and Fig. \ref{Figdelta} (c) for fixed $a$. These figures display that the resulting shadow is smaller (larger) than the Schwarzschild BH shadow for BHs located in a low (high) curvature background. Fig. \ref{Figdelta} (d) indicates the 
difference between the obtained shadow and the Schwarzschild shadow diameter, in $ (a,Q) $ plane for fixed $ \Lambda $ in the AdS spacetime, revealing that only electric charges in the range $Q<0.85$ can satisfy the mentioned constraint. The variation of $\delta$ with the change in parameter $ a $ and $ \Lambda $ is depicted in Fig. \ref{Figdelta} (e) for a fixed electric charge which indicates that the constraint on $\delta$ is satisfied for the range $-0.044<\Lambda <0$. According to this figure, the shadow cast by the EEH-AdS/dS BH is larger than the Schwarzschild shadow for small $ \vert \Lambda \vert $, and vice versa. In Fig. \ref{Figdelta} (f), the variation of $\delta$ is shown in $ (Q,\Lambda) $ plane for fixed $ a $, illustrating that the resulting shadow is larger than the Schwarzschild BH shadow for small $ \vert \Lambda \vert $.  Looking closely at Fig. \ref{Figdelta}, one can notice that the influence of the EH parameter on $\delta$ is negligible compared to the other two parameters. 

	\subsection{Energy emission rate}
	\label{rate}


In this subsection, we examine how the parameters of the theory affect the energy emission from the BH solution (\ref{a11}). The thermal radiation emitted by a BH is directly related to its Hawking temperature. According to quantum field theory in curved spacetime, BH radiation arises from the spontaneous creation of virtual particle-antiparticle pairs near the event horizon. In this process, the antiparticle with negative energy falls into the BH, while the particle with positive energy escapes to infinity via quantum tunneling. As a result, the BH loses mass over time, a phenomenon known as Hawking radiation \cite{Hawking:1975}.

At high energies, the BH's absorption cross-section plays a significant role in determining its radiation profile. It has been shown that, for distant observers, the BH shadow is closely related to its high-energy absorption cross-section \cite{Sanchez:a,Decanini:a}. Specifically, the absorption cross-section tends to oscillate around a limiting constant value (denoted by 
$ \sigma_{lim}$) as the energy increases. This limiting value is approximately equal to the geometrical cross-section of the photon sphere, i.e., 
$\sigma_{lim}\approx r_{sh}^{2}  $.  The energy emission rate is expressed as \cite{Wei:2013}
\begin{equation}
\frac{d^{2}\mathcal{E}(\omega )}{dtd\omega }=\frac{2\pi ^{3}\omega
^{3}r_{sh}^{2}}{e^{\frac{\omega }{T}}-1},  \label{Eqemission}
\end{equation}%
in which $\omega $ represents the emission frequency and $T$ denotes the Hawking temperature calculated as
	\begin{equation}
		T=\frac{1}{4 \pi r_{eh}}\left(1-\frac{Q^{2}}{r_{eh}^{2}}+\frac{a Q^{4}}{4 r_{eh}^{6}}-\Lambda r_{eh}^{2}\right).
	\end{equation}
	
\begin{figure}[!htb]
		\centering
		\subfloat[ $ Q=0.9 $ and $ \Lambda=0.02$]{
			\includegraphics[width=0.31\textwidth]{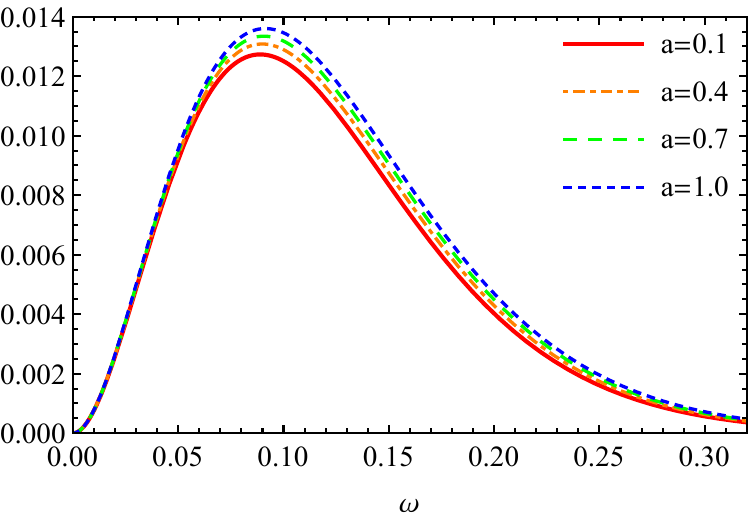}}
		\subfloat[$ a=0.2 $ and $ \Lambda=0.02$]{
			\includegraphics[width=0.31\textwidth]{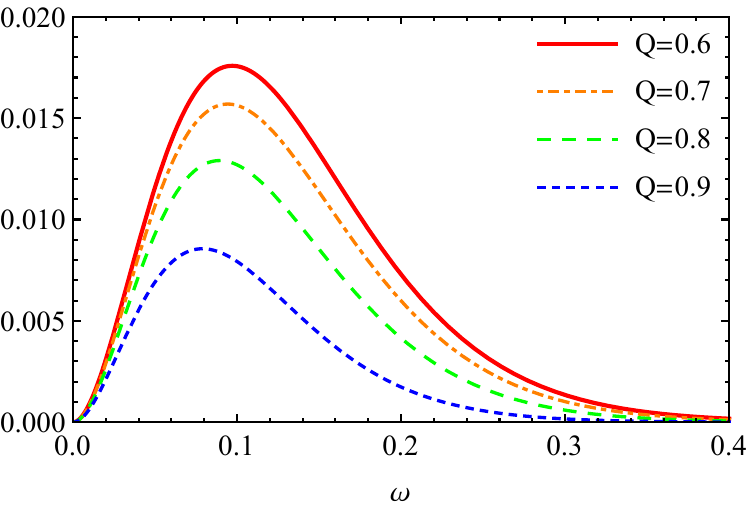}}       
		\newline
		\subfloat[$ a=0.2 $ and $ Q=0.8$]{
			\includegraphics[width=0.31\textwidth]{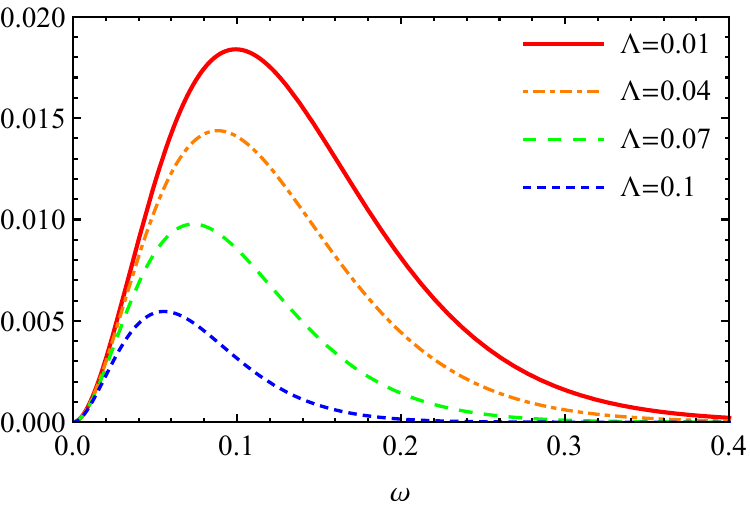}}
		\subfloat[$ a=0.2 $ and $ Q=0.8$]{
			\includegraphics[width=0.31\textwidth]{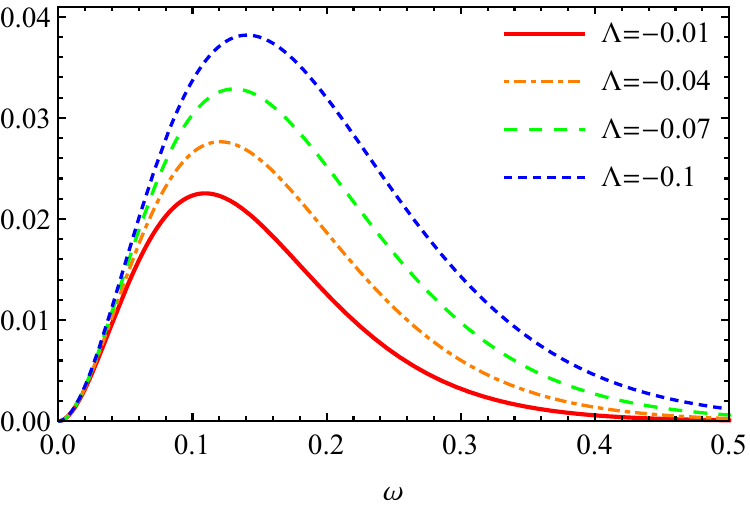}}        
		\newline
		\caption{The energy emission rates for the EEH-AdS/dS BH with different values of parameters $ Q $, $ a $ and $ \Lambda $.}
		\label{FigEr}
	\end{figure}

To show how the energy emission rate is affected by the parameters of the model, we have plotted Fig \ref{FigEr}. It can be seen from Fig. \ref{FigEr}(a) that the energy emission rate increases with an increase in the EH parameter, meaning that the evaporation process would be faster when the effect of this parameter gets stronger. This reveals the fact that these BHs have shorter lifetimes compared to BHs in Maxwell's theory. According to Fig. \ref{FigEr}(b), the electric charge effect on the emission rate is opposite to that of the EH parameter,  representing that electrically charged BHs have longer lifetimes than neutral BHs. To study the effect of the cosmological constant, we plotted Figs. \ref{FigEr}(c) and \ref{FigEr}(d) which show that the radiation rate will be low in a high curvature background in dS space-time. While in AdS space-time, increasing the absolute value of the cosmological constant leads to a fast emission of particles (see Fig. \ref{FigEr}(d)). This shows that BHs have shorter lifetimes in AdS spacetime compared to BHs in dS spacetime.

\section{GAUSS-BONNET THEOREM AND DEFLECTION ANGLE}
\label{def}
\begin{figure}[!htb]
		\centering
	{\includegraphics[width=0.40\textwidth]{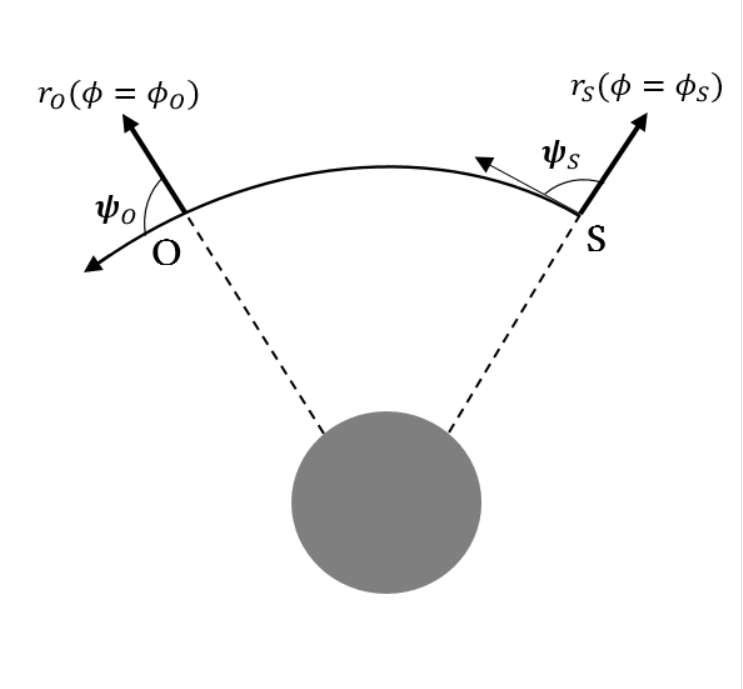}}        
		\newline
		\caption{Schematic diagram illustrating light propagation in a spherically symmetric spacetime. The black hole is located at the origin $ r=0 $. The angles 
$\phi_{O}$ and $\phi_{S}$
  represent the angles between the light ray and the radial direction at the observer and source, respectively. The coordinate angle $ \phi_{OS} $ represents the angular separation between source and observer as seen from the center.}
		\label{Figdef0}
	\end{figure}

The main objective of this section is to study the gravitational deflection of light around the EEH-AdS/dS BH.  Gibbons and Werner have recently introduced a geometric approach to calculating gravitational deflection angle in the weak-field range, known as the Gauss-Bonnet method \cite{Gibbons:235,Gibbons:3047}. The Gauss-Bonnet theorem (GBT) offers an elegant and global perspective on the deflection problem by representing the deflection angle as an integral of curvature. This approach, which relates the effect of distortion to the curvature and topology of space, has gained popularity among researchers in recent years. Ishihara et al. used GBT to study the gravitational deflection of light with the
observer and source at finite distance in both the weak and strong deflection limits \cite{Ishihara:abc,Ishihara:def}. The finite-distance deflection aligns more closely with our physical world because it assumes that both the observer and source are at finite distances from a gravitational lens. Furthermore, for cases with non-flat asymptotically spacetimes, the asymptotic deflection angle will diverge, but finite-distance deflection angles can exist. Accordingly, we employ GBT to calculate the finite-distance deflection angle for the EEH-AdS/dS BH. 	The deflection angle can be defined as \cite{Kumaran:ao,Takizawa:101}
	\begin{equation}\label{a30}
	\alpha=\psi_{O}-\psi_{S}+\phi_{OS},
	\end{equation}
where $\phi_{OS}\equiv$ $\phi_{O}-\phi_{S}$ represents the angular separation between the observer (receiver) and the source. The quantities $\phi_{O}$ and $\phi_{S}$ denote the angular (or coordinate) positions of the observer and source relative to the lens. Additionally, $\psi_{O}$ and $\psi_{S}$ are the angles measured by the observer and source, respectively, with respect to the radial direction (see Fig. \ref{Figdef0}). These angles are defined as follows:
 \cite{Ishihara:abc,Jafarzade:kz}
	\begin{equation}\label{a31}
		\sin\psi=\sqrt{\frac{H(r)f(r)}{r^{2}h(r)}}.
	\end{equation}
	
Due to the spherically symmetric property of
the BH under study, one can assume that the deflection of light occurs in the equatorial plane ($\theta=\frac{\pi}{2}$). Using the photon orbital equation in relation $\eqref{1-16}$ and the inverse definition of $r=u^{-1}$, we have
\begin{equation}\label{a34}
	\left(\frac{du}{d\phi}\right)^{2}=G(u),
\end{equation}
where $ G(u) $ defined in Eq. $\eqref{Gu1}$ is calculated as 
\begin{equation}\label{Gu2}
	G(u)=\frac{1}{b^{2}}+\frac{1}{3}\Lambda -u(\phi)^{2}+2Mu(\phi)^{3}-Q^{2}u(\phi)^{4}-\frac{1}{3}\Lambda Q^{2}u(\phi)^{4}-\frac{2a}{b^{2}}u(\phi)^{4}+a Q^{2}u(\phi)^{6}-2aMQ^{2}u(\phi)^{7}.
\end{equation}

From the above equation, $u(\phi)  $ is obtained as
\begin{equation}\label{a340}
	u(\phi)=\frac{\sin\phi}{b}+Mu_{1}(\phi)+\Lambda u_{2}(\phi)+Q^{2}u_{3}(\phi)+aQ^{2}u_{4}(\phi)+aMQ^{2}u_{5}(\phi)+a\Lambda Q^{2}u_{6}(\phi),
\end{equation}
in which
\begin{eqnarray}
u_{1}(\phi)&=&\frac{1+\cos(\phi)^{2}}{b^{2}},\\ \nonumber
u_{2}(\phi)&=&\frac{1}{6}b \sin(\phi) ,\\ \nonumber
u_{3}(\phi)&=&\frac{6\phi\cos(\phi)-4\sin(\phi) -\cos(\phi) \sin(2\phi)}{8b^{3}},\\ \nonumber
u_{4}(\phi)&=&\frac{36\phi\cos(\phi)-32\sin(\phi) -\cos(\phi) \sin(4\phi)}{64b^{4}},\\ \nonumber
u_{5}(\phi)&=&-\frac{5040-1920\cos(\phi)+3145\cos(2\phi)+16\cos(4\phi)-9\cos(6\phi)+1080\phi\sin(2\phi)}{1920b^{5}}\\ \nonumber
&-& \frac{240-240\cos(\phi)+5\cos(2\phi)+8\cos(4\phi)+3\cos(6\phi)}{120b^{6}},\\ \nonumber
u_{6}(\phi)&=&\frac{32\sin(\phi)-36\phi\cos(\phi)+\cos(\phi)\sin(4\phi)}{384b^{2}}-\frac{64\sin(\phi)-\cos(\phi)\left(60\phi +8\sin(2\phi)-3\sin(4\phi) \right) }{192b^{3}}.\\ \nonumber
\end{eqnarray}

To calculate the deflection angle, we need an explicit expression for $\phi_{OS}$ \cite{Kumaran:ao}
	\begin{equation}\label{a32}
		\phi_{OS}=\int_{S}^{O}d\phi=\int_{u_{S}}^{u_{0}}\left(\frac{du}{d\phi}\right)^{-2}du-\int_{u_{0}}^{u_{O}}\left(\frac{du}{d\phi}\right)^{-2}du.
	\end{equation}
	
	Now, by combining relations $\eqref{a30}$ and $\eqref{a32}$, the deflection angle is obtained as
	\begin{equation}\label{a33}
		\alpha=\int_{u_{S}}^{u_{0}}\left(\frac{du}{d\phi}\right)^{-2}du-\int_{u_{0}}^{u_{O}}\left(\frac{du}{d\phi}\right)^{-2}du+\psi_{O}-\psi_{S},
	\end{equation}
To obtain the values $\psi_{O}$ and $\psi_{S}$, we use equation $\eqref{a31}$ which leads to  
	\begin{align}\label{a36}
		\psi_{O}-\psi_{S}&=\left( \sin^{-1}(bu_{O})+\sin^{-1}(bu_{S})-\pi\right) -bM\left(\frac{u_{O}^{2}}{\sqrt{1-b^{2}u_{O}^{2}}}+\frac{u_{S}^{2}}{\sqrt{1-b^{2}u_{S}^{2}}}\right) \nonumber\\
	&-\frac{b \Lambda}{6}\left(\frac{u_{O}^{-1}}{\sqrt{1-b^{2}u_{O}^{2}}}+\frac{u_{S}^{-1}}{\sqrt{1-b^{2}u_{S}^{2}}}\right)+\frac{M\Lambda b^{3}}{6}\left(\frac{u_{O}^{2}}{(1-b^{2}u_{O})^{\frac{3}{2}}}+\frac{u_{S}^{2}}{(1-b^{2}u_{S})^{\frac{3}{2}}}\right) \nonumber\\
	&-\frac{bM\Lambda}{6}\left(\frac{1}{\sqrt{1-b^{2}u_{O}^{2}}}+\frac{1}{\sqrt{1-b^{2}u_{S}^{2}}}\right)+\left(-\frac{1}{24}ba\Lambda Q^{2}+\frac{b Q^{2}}{2}\right)\left(\frac{u_{O}^{3}}{\sqrt{1-b^{2}u_{O}^{2}}}+\frac{u_{S}^{3}}{\sqrt{1-b^{2}u_{S}^{2}}}\right)\nonumber\\
	&+\frac{1}{12}b\Lambda Q^{2} \left(\frac{u_{O}}{\sqrt{1-b^{2}u_{O}^{2}}}+\frac{u_{S}}{\sqrt{1-b^{2}u_{S}^{2}}}\right)-\frac{1}{12}b^{3}Q^{2}\Lambda \left(\frac{u_{O}^{3}}{(1-b^{2}u_{O}^{2})^{\frac{3}{2}}}+\frac{u_{S}^{3}}{(1-b^{2}u_{S}^{2})^{\frac{3}{2}}}\right)\nonumber\\
	&-\frac{1}{2}b^{3}Q^{2}M \left(\frac{u_{O}^{6}}{(1-b^{2}u_{O}^{2})^{\frac{3}{2}}}+\frac{u_{S}^{6}}{(1-b^{2}u_{S}^{2})^{\frac{3}{2}}}\right)+\frac{1}{2}bMQ^{2}\left(\frac{u_{O}^{4}}{\sqrt{1-b^{2}u_{O}^{2}}}+\frac{u_{S}^{4}}{\sqrt{1-b^{2}u_{S}^{2}}}\right)\nonumber\\
	&+\frac{1}{4}abQ^{2}\left(\frac{u_{O}^{5}}{\sqrt{1-b^{2}u_{O}^{2}}}+\frac{u_{S}^{5}}{\sqrt{1-b^{2}u_{S}^{2}}}\right)-\frac{1}{4}abMQ^{2} \left(\frac{u_{O}^{6}}{\sqrt{1-b^{2}u_{O}^{2}}}+\frac{u_{S}^{6}}{\sqrt{1-b^{2}u_{S}^{2}}}\right)\nonumber\\
	&-\frac{1}{4}a M b^{3}Q^{2} \left(\frac{u_{O}^{8}}{(1-b^{2}u_{O}^{2})^{\frac{3}{2}}}+\frac{u_{S}^{8}}{(1-b^{2}u_{S}^{2})^{\frac{3}{2}}}\right)-\frac{1}{24}a \Lambda b^{3}Q^{2} \left(\frac{u_{O}^{5}}{(1-b^{2}u_{O}^{2})^{\frac{3}{2}}}+\frac{u_{S}^{5}}{(1-b^{2}u_{S}^{2})^{\frac{3}{2}}}\right).
	\end{align}

Then for $\phi_{OS}$, using Eq. $\eqref{a34}$, $\eqref{Gu2}$ and $\eqref{a340}$ and the relation $\eqref{a32}$, we have 
	\begin{align}\label{a37}
		\phi_{OS}&=-\left( \sin^{-1}(bu_{O})+\sin^{-1}(bu_{S})-\pi\right) +\frac{M}{b}.\left(\frac{2-b^{2}u_{0}^{2}}{\sqrt{1-b^{2}u_{O}^{2}}}+\frac{u_{S}}{\sqrt{1-b^{2}u_{S}^{2}}}\right)+\frac{b^{3}\Lambda}{6}\left(\frac{u_{O}}{\sqrt{1-b^{2}u_{O}^{2}}}+\frac{u_{S}}{\sqrt{1-b^{2}u_{S}^{2}}}\right) \nonumber\\
		&-\frac{4bQ^{2}+3aQ^{2}}{16b}.\left(\frac{u_{O}(3-b^{2}u_{O}^{2})}{\sqrt{1-b^{2}u_{O}^{2}}}+\frac{u_{S}(3-b^{2}u_{S}^{2})}{\sqrt{1-b^{2}u_{S}^{2}}}\right)+\frac{3 Q^{2}(3a+4b)}{16 b^{2}}\left(\sin^{-1}(bu_{O})+\sin^{-1}(bu_{S})\right) \nonumber\\
			&+\frac{M\Lambda b}{6}\left(\frac{2-3b^{2}u_{O}^{2}}{(1-b^{2}u_{O}^{2})^{\frac{3}{2}}}+\frac{2-3b^{2}u_{S}^{2}}{(1-b^{2}u_{S}^{2})^{\frac{3}{2}}}\right)+\frac{1}{48}\Lambda a Q^{2}(5-3b)\left(\frac{u_{O}(3-4b^{2}u_{O}^{2})}{(1-b^{2}u_{O}^{2})}^{\frac{3}{2}}+\frac{u_{S}(3-4b^{2}u_{S}^{2})}{(1-b^{2}u_{S}^{2})^{3}{2}}\right)\nonumber\\
			&-\frac{3MQ^{2}}{2b}\left(\frac{u_{O}(1+2bu_{R}-3b^{2}u_{O}^{2})}{(1-b^{2}u_{O}^{2})^{\frac{3}{2}}}+\frac{u_{S}(1+2bu_{S}-3b^{2}u_{S}^{2})}{(1-b^{2}u_{S}^{2})^{\frac{3}{2}}}\right)\nonumber\\
		&-\frac{3MQ^{2}}{2b^{2}}\left(\frac{\sin^{-1}(bu_{O})(1+bu_{O}-3b^{2}u_{O}^{2})}{(1-b^{2}u_{O}^{2})^{2}}+\frac{\sin^{-1}(bu_{S})(1-2b^{2}u_{S}^{2})}{(1-b^{2}u_{S}^{2})^{2}}\right)\nonumber\\	
	&-\frac{1}{8}\Lambda b Q^{2}\left(\frac{\sin^{-1}(bu_{O})(1-2b^{2}u_{O}^{2})}{(1-b^{2}u_{O}^{2})^{2}}+\frac{\sin^{-1}(bu_{S})(1-2b^{2}u_{S}^{2})}{(1-b^{2}u_{S}^{2})^{2}}\right)\nonumber\\
&+\frac{1}{16b}\Lambda a Q^{2}\left(\frac{\sin^{-1}(bu_{O})(5-3b-10b^{2}u_{O}^{2})}{(1-b^{2}u_{O}^{2})^{2}}+\frac{\sin^{-1}(bu_{S})(5-3b-10b^{2}u_{S}^{2})}{(1-b^{2}u_{S}^{2})^{2}}\right)\nonumber\\
&+\frac{MaQ^{2}(2+b)}{b^{4}}-\frac{MaQ^{2}}{24b^{4}}\left(\frac{(49+4b-6b^{3}u_{O}^{2})}{(-b^{2}u_{O}^{2}+1)^{\frac{3}{2}}}+\frac{(49+4b-6b^{3}u_{S}^{2})}{(-b^{2}u_{S}^{2}+1)^{\frac{3}{2}}}\right).
\end{align}

\begin{figure}[!htb]
\centering
\subfloat[ $ Q=0.8 $ and $ \Lambda=0.02$]{
\includegraphics[width=0.31\textwidth]{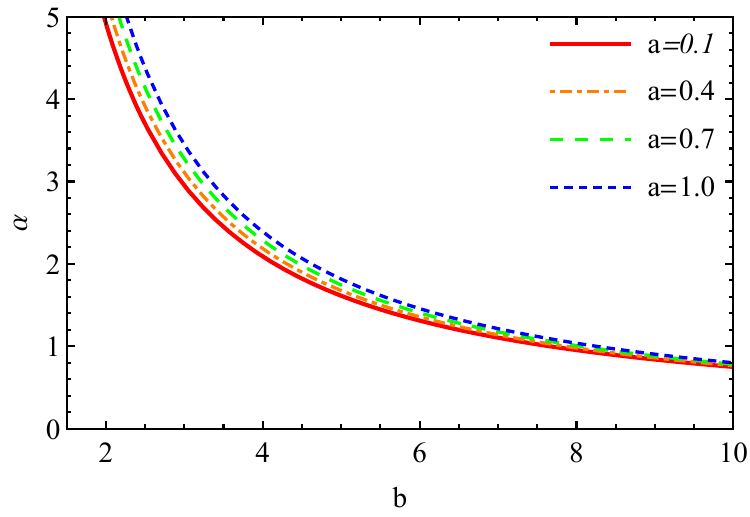}}
\subfloat[$ a=0.2 $ and $ \Lambda=0.02$]{
     \includegraphics[width=0.31\textwidth]{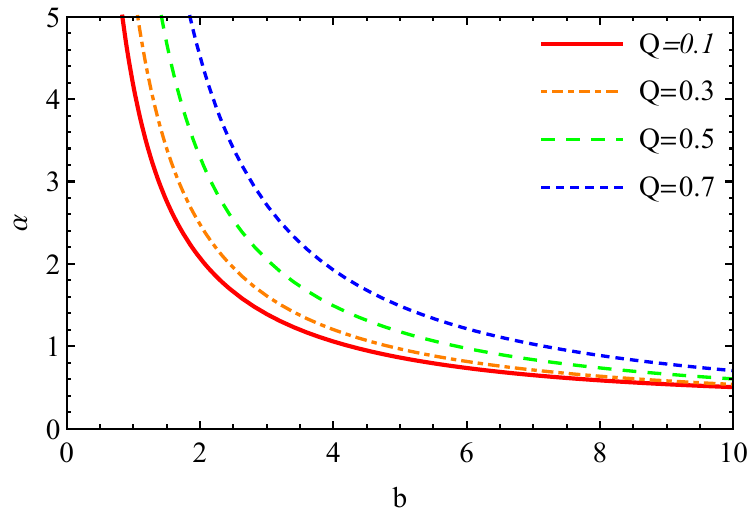}}       
\newline
\subfloat[$ a=0.2 $ and $ Q=0.8$]{
\includegraphics[width=0.31\textwidth]{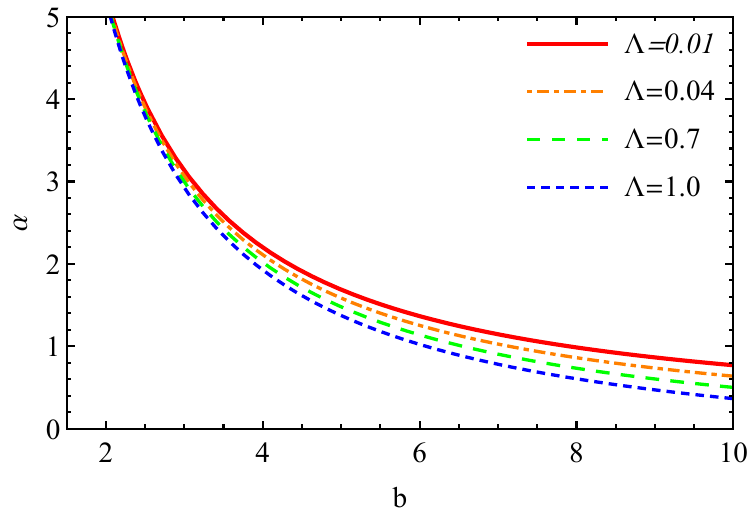}}
\subfloat[$ a=0.2 $ and $ Q=0.8$]{
     \includegraphics[width=0.31\textwidth]{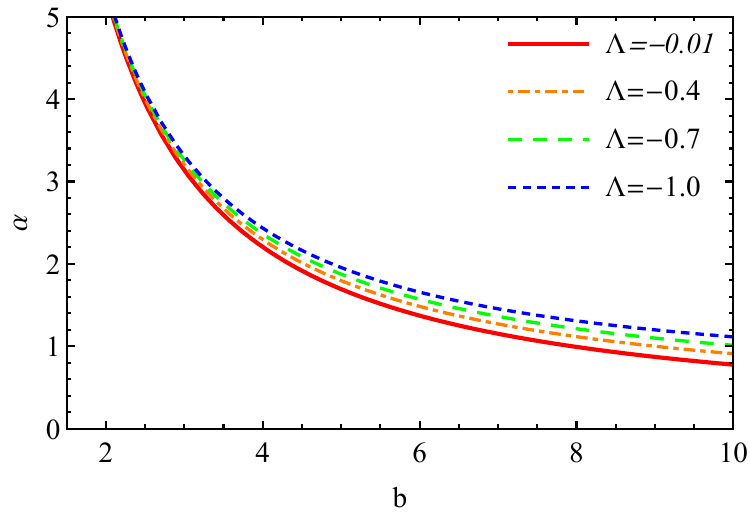}}        
\newline
\caption{The behavior of $ \alpha $ with respect to the impact parameter $ b $ for $ M=1 $ and different values of parameters $ Q $, $ a $ and $ \Lambda $.}
\label{Figdef}
\end{figure}

Inserting $\eqref{a36}$ and $\eqref{a37}$ in $\eqref{a33}$, one can obtain an expression for the deflection angle. Assuming $u_{O}\to 0$ and  $u_{S}\to 0$, one can calculate the deflection angle at the infinite distance limit. Fig. $\eqref{Figdef}$ displays how $ \alpha $ changes under varying the parameters of the theory. In Fig. \ref{Figdef}(a), we set the electric charge and cosmological constant as fixed parameters and examined the influence of the EH parameter on $ \alpha $. From this figure, we notice that increasing the parameter $ a $ leads to increasing the deflection angle.
 Fig. \ref{Figdef}(b) illustrates the electric charge effect on $ \alpha $ for fixed $ a $ and $ \Lambda $, indicating that this parameter has an increasing effect on the deflection angle like the EH nonlinear field. This reveals that photons are more deflected from their straight path around electrically charged BHs.
 To study the impact of the curvature background on $ \alpha $, we plotted Figs. \ref{Figdef}(c) and \ref{Figdef}(d) while keeping fixed values for the electric charge and EH parameter. As we see, the effect of $\Lambda  $ will be significant for large impact parameters. 
 In dS spacetime, the cosmological constant has a decreasing contribution on  $ \alpha $ (see Fig. \ref{Figdef}(c)),  whereas in AdS spacetime, the effect of $\Lambda  $ is to increase the deflection angle
(see Fig. \ref{Figdef}(d)). As a result, the light rays will be more deflected on a background with low (high) curvature in dS (AdS) spacetime.

	\section{Conclusion}
	\label{conclusion}
	
	Recent observations of the supermassive BHs $ M87^{*} $ and Sgr A$^{*} $ by the EHT have created a wide range of new scientific opportunities in gravitational physics, compact objects, and relativistic astrophysics. Therefore,  studying the optical appearance of BHs is a fascinating topic in astrophysics,  playing a crucial role in several high-energy situations. Motivated by this, we 
performed a study in the context of the optical features of EEH-AdS/dS BHs and investigated how the parameters of the model affect them. 
	
	As a first attempt,  the trajectories of light rays categorized as direct, lensed, and photon rings on the equatorial plane have been studied, and the effects of the electric charge and cosmological constant on the range of lensing and photon rings has been examined. According to our findings, the EH parameter has a negligible effect on the trajectories of photons, while both the electric charge and cosmological constant have a remarkable influence on the classification of light trajectories. So that with increasing (decreasing) the electric charge (cosmological constant), the range of lensing and photon rings increases.
	
	In the next step, a study on the size and geometrical shape of the shadow for the associated BHs was presented. To have an acceptable optical behavior, we found the allowed range of parameters for which the constraints $r_{eh}<r_{ph}< r_{sh}$ are satisfied. Then we studied the influence of the parameters on the shadow size and noticed that both the EH parameter and the electric charge decrease the radius of the BH shadow. Regarding the effect of the cosmological constant, our findings showed that the cosmological constant has an increasing contribution to the shadow radius in dS spacetime, while in AdS spacetime, the shadow size decreases with an increase in the absolute value of the cosmological constant. 
	
	Additionally, using the latest observational data from EHT on $ M87^{*} $, we have provided constraints on the parameters of EEH-AdS/dS BHs and found that EEH-dS BHs are a suitable candidate for the supermassive BH $M87^{*}$ compared to EEH-AdS BHs. We also extracted bounds on the parameters by studying the Schwarzschild shadow deviation ($ \delta $), which measures the difference between the model shadow diameter and the Schwarzschild BH shadow diameter. Based on our analysis,  the constraint on $ \delta $ will be violated for small values of $Q$ in dS spacetime, while in AdS spacetime, the constraint will be violated for a large electric charge. Moreover, we observed that the shadow size of the EEH-dS BH (EEH-AdS BH) becomes smaller than the shadow radius of the Schwarzschild BH for an intermediate  (small) electric charge. Regarding the influence of $\Lambda$ on the shadow deviation ($ \delta $), we found that the resulting shadow is larger (smaller) than the Schwarzschild BH shadow for BHs located in a high (low) curvature background in dS spacetime. While the opposite behavior is observed in AdS spacetime.
	
	We extended our study by investigating the energy emission rate around such BHs and found that the evaporation process becomes faster with increasing the EH parameter, which shows that these BHs have shorter lifetimes in the presence of the EH nonlinear field. Examining the behavior of the energy emission rate with changing the electric charge revealed that electrically charged BHs have longer lifetimes than neutral BHs. Additionally, we investigated the effect of the cosmological constant on the energy emission rate and noticed that the evaporation process will be slower in a high curvature background in dS spacetime, while in AdS spacetime, increasing $ \vert \Lambda \vert $ leads to a fast emission of particles. As a result, EEH-AdS BHs have shorter lifetimes than EEH-dS BHs.
	
Finally, a study of the gravitational lensing of light around such BHs was performed in the weak-field approximation. Considering the observer and source at finite distances from the EEH-AdS/dS BHs, we derived an analytical formula for the deflection angle. Our analysis of the influence of parameters on the deflection angle of photons showed that both the EH parameter and the electric charge make an increasing contribution to the deflection angle. In other words, the deviation of photons from the straight path will be significant in the presence of EH nonlinear and electric fields. By studying the effect of the cosmological constant  on the deflection angle $ \alpha $, we observed that increasing $\Lambda $ leads to a decrease in $ \alpha $ in dS spacetime, while in AdS spacetime, $ \alpha $ increases with increasing $\vert \Lambda \vert $. This indicates that light rays are more deflected in a background with low (high) curvature in dS (AdS) spacetime.


\begin{thebibliography}{99}


	\bibitem{EventHorizonTelescope:2019dse}
		K. Akiyama et al. [Event Horizon Telescope], Astrophys. J. Lett. \textbf{875}, L1 (2019).
	  \bibitem{EventHorizonTelescope:2019uob}
		K. Akiyama et al. [Event Horizon Telescope], Astrophys. J. Lett. \textbf{875},  L2 (2019). 
		
		\bibitem{EventHorizonTelescope:2019jan}
		K. Akiyama et al. [Event Horizon Telescope],	Astrophys. J. Lett. \textbf{875},  L3 (2019).
		\bibitem{EventHorizonTelescope:2019ths}
		K. Akiyama et al. [Event Horizon Telescope],
		Astrophys. J. Lett. \textbf{875},  L4 (2019).
		\bibitem{EventHorizonTelescope:2019pgp}
		K. Akiyama et al. [Event Horizon Telescope],
		Astrophys. J. Lett. \textbf{875},  L5 (2019).
		\bibitem{EventHorizonTelescope:2019ggy}
		K. Akiyama et al. [Event Horizon Telescope],
		Astrophys. J. Lett. \textbf{875},  L6 (2019).
		
\bibitem{Psaltis:51}		
D. Psaltis, Gen. Relativ. Gravit. \textbf{51},  137 (2019).
\bibitem{Dokuchaev:583}
V. I. Dokuchaev, N. O. Nazarova, Physics-Uspekhi \textbf{63}, 583 (2020).

\bibitem{Chael:918}
A. Chael, M. D. Johnson and A. Lupsasca, Astrophys. J. \textbf{918}, 6 (2021).



	
\bibitem{Vries:1999123}
A. de Vries, Class. Quantum Gravit. 17, 123 (1999).
 
\bibitem{Johannsen:446}
T. Johannsen, and D. Psaltis, Astrophys. J. 718, 446  (2010).
 
\bibitem{Cunha:024039}
 P. V. P. Cunha, C. A. R. Herdeiro, and E. Radu, Phys. Rev. D 96, 024039 (2017).		
		
\bibitem{Pedro:5042}
 P. V. P. Cunha, and C. A. R. Herdeiro, Gen. Rel. Grav. 50, 42  (2018).

\bibitem{Wei:08030}
 S. -W. Wei, Y. -C. Zou, Y. -X. Liu, and R. B. Mann, J. Cosmol.
 Astropart. Phys. 08, 030  (2019).

\bibitem{Belhaj:215004}
A. Belhaj, M. Benali, A. El Balali, H. El Moumni, and S. E. Ennadifi, Class. Quantum Gravit. 37, 215004 (2020).		
		
\bibitem{Synge:1966okc}
J. L. Synge, Mon. Not. Roy. Astron. Soc. \textbf{131}, 463 (1966).		
		
\bibitem{Carter:1968rr}
	B.~Carter, Phys. Rev. \textbf{174}, 1559 (1968).
	
\bibitem{Bardeen:1972fi}
J. M. Bardeen, W. H. Press and S. A. Teukolsky, Astrophys. J. \textbf{178}, 347 (1972).	


\bibitem{Liu:858}
 W. Liu, D. Wu, J. Wang, Phys. Lett. B \textbf{858} 139052 (2024).
 
 \bibitem{Wang:05}
 W. Liu, D. Wu, J. Wang, JCAP \textbf{05}, 017 (2025).
 
 \bibitem{Fang:08}
 W. Liu, D. Wu, X. Fang, J. Jing, J. Wang, JCAP \textbf{08}, 035 (2024).


\bibitem{Vagnozzi:40}
 S. Vagnozzi et al., Class. Quantum Grav. \textbf{40},
165007 (2023).


\bibitem{Khodadi:26932}
M. Khodadi, S. Vagnozzi, and J. T. Firouzjaee, Sci.
Rep. \textbf{14}, 26932 (2024).

\bibitem{Bambi:100}
C. Bambi, K. Freese, S. Vagnozzi, and L. Visinelli, Phys.
Rev. D \textbf{100}, 044057 (2019).


 
 \bibitem{Jafarzade:1a}
 K. Jafarzade, M. Ghasemi-Nodehi, F. Sadeghi, B. Mirza, JCAP \textbf{03}, 063 (2025).
 
 \bibitem{Jafarzade:1b}
 K. Jafarzade, S. Shaymatov, M. Jamil, Astropart. Phys \textbf{168},  103100 (2025).
 
\bibitem{Zheng:1d}
H, Zheng, M. Wu, G-P. Li, Q. Jiang, Eur. Phys. J. C \textbf{85}, 46 (2025).
\bibitem{Gao:1d}
X-J. Gao, Eur. Phys. J. C \textbf{84}, 973 (2024).

\bibitem{Erices:1a}
C. Erices, M. Fathi,  JCAP \textbf{01}, 016 (2025).


 
 
 



\bibitem{Feng:73}
H. Feng, R-J. Yang, W-Q. Chen, Astropart. Phys \textbf{166}, 103075 (2025).


\bibitem{Kumar:44}
S. Kumar, A. Uniyal, S. Chakrabarti, Phys. Dark Univ. \textbf{44},  101472 (2024).

\bibitem{Maqsood:047}
M. Ali Raza, M. Zubair, E. Maqsood, JCAP \textbf{05}, 047 (2024).


\bibitem{Atamurotov:2023}
F. Atamurotov, M. Jamil, K. Jusufi, 	Chin. Phys. C  \textbf{47},  035106  (2023).

\bibitem{Briozzo:2023}
G. Briozzo, E. Gallo, T. M\"{a}dler, Phys. Rev. D \textbf{107}, 124004 (2023).


\bibitem{Bakopoulos:110}
 A. Bakopoulos, T. Karakasis, N. E. Mavromatos,
T. Nakas, and E. Papantonopoulos, Phys. Rev. D
\textbf{110}, 024014 (2024).

\bibitem{Zhong:103}
 Z. Hu, Z. Zhong, P. C. Li, M. Guo and B. Chen,Phys. Rev. D \textbf{103}, 044057 (2021).

\bibitem{Zhong:104}
 Z. Zhong, Z. Hu, H. Yan, M. Guo and B. Chen, Phys. Rev. D \textbf{104}, 104028 (2021).

\bibitem{Aliyan:1b}
F. Aliyan, K. Nozari, Phys. Dark. Univ.  \textbf{46},  101611 (2024).

\bibitem{Zeng:764}
X. Zeng, K-J. He, G-P. Li, E-W. Liang, S. Guo, Eur. Phys. J. C \textbf{82}, 764 (2022).

\bibitem{Guzman:11}
E. Guzman-Herrera, A. Montiel, N. Breton,  JCAP \textbf{11}, 002 (2024).

\bibitem{Lambiase:48}
G. Lambiase, D. Jyoti Gogoi, R. C. Pantig, A. \"{O}vg\"{u}n, Phys. Dark Univ. \textbf{48}, 101886 (2025).

\bibitem{Hamil:73}
B. Hamil, B. C. L\"{u}tf\"{u}oglu, Forschr. Phys. \textbf{73}, 2400105 (2025).




\bibitem{Born:1934gh}
M. Born and L. Infeld, Proc. Roy. Soc. Lond. A \textbf{144}, 852 (1934).

\bibitem{Heisenberg:1936nmg}
W. Heisenberg and H. Euler, Z. Phys. \textbf{98}, 714 (1936).


\bibitem{Magnea:2004ai}
L. Magnea, R. Russo and S. Sciuto, Int. J. Mod. Phys. A \textbf{21}, 533 (2006).

\bibitem{Sciuto:2005sq}
S. Sciuto, Afr. J. Math. Phys. \textbf{3}, 45 (2006).

\bibitem{Novello:1999pg}
M. Novello, V. A. De Lorenci, J. M. Salim and R. Klippert, Phys. Rev. D \textbf{61}, 045001 (2000).
	
	\bibitem{ATLAS:2017fur}
M. Aaboud et al.,
Nature Phys. \textbf{13}, 852 (2017).


\bibitem{ATLAS:2019azn}
G. Aad et al.,
Phys. Rev. Lett. \textbf{123}, 052001 (2019).

\bibitem{Capparelli:2017mlv}
L. M. Capparelli, A. Damiano, L. Maiani and A. D. Polosa, Eur. Phys. J. C \textbf{77},  754 (2017).
	
		
\bibitem{Schwinger:1951nm}
J. S. Schwinger, Phys. Rev. \textbf{82}, 664 (1951).

\bibitem{Euler:1935zz}
H. Euler and B. Kockel, Naturwiss. \textbf{23}, 246 (1935).


\bibitem{Euler:1935qgl}
H. Euler, Annalen Phys. \textbf{26},  398 (1936).


\bibitem{Adler:1970gg}
S. L. Adler, J. N. Bahcall, C. G. Callan and M. N. Rosenbluth,
Phys. Rev. Lett. \textbf{25}, 1061 (1970).
			
\bibitem{adler1}
S. L. Adler, Annals Phys. (N.Y) \textbf{67}, 599 (1971).


\bibitem{Okyay:2021nnh}
M. Okyay and A. \"{O}vg\"{u}n, JCAP \textbf{01},  009 (2022).
		
\bibitem{Dittrich:1998fy}
	W. Dittrich and H. Gies,
	Phys. Rev. D \textbf{58}, 025004 (1998).
		
\bibitem{Plebanski:1970zz}
	J. Plebanski, \textit{Lectures on nonlinear electrodynamics}, RX-476.
	
\bibitem{Theodosopoulos:84}
D. P. Theodosopoulos, T. Karakasis, G. Koutsoumbas, E. Papantonopoulos, Eur. Phys. J. C \textbf{84}, 592 (2024).
		
\bibitem{Allahyari:1a}
 A. Allahyari, M. Khodadi, S. Vagnozzi, D.F. Mota, J. Cosmol. Astropart. Phys.
2002 (2020) 003.
		
\bibitem{Zeng:2022pvb}
		X. X. Zeng, K. J. He, G. P. Li, E. W. Liang and S. Guo,
		Eur. Phys. J. C \textbf{82},  764 (2022).

\bibitem{Kim:2022xum}
J. Y. Kim, Eur. Phys. J. C \textbf{82},  485 (2022).




\bibitem{Cheng:yh}
Y. Zhao, H. Cheng, Chinese Physics C \textbf{48}, 125106 (2024). 

\bibitem{Gursel:hm}
H. Gursel, M. Mangut, E. Sucu, arXiv:2503.12306. 

\bibitem{Qiao:su}
P. Su, C-K. Qiao, arXiv:2410.02411.

\bibitem{Jiang:yh}
 Y-H. Jiang, T. Wang, Phys. Rev. D \textbf{110}, 103009 (2024).
 
 \bibitem{Lambiase:gd}
 G. Lambiase, D. J. Gogoi, R. C. Pantig, A.  \"Ovg\"un, Phys. Dark Univ. \textbf{48}, 101886 (2025).
 \bibitem{Myung:cv}
 Y. S. Myung, arXiv:2503.18239.
 
 \bibitem{Guzman:an}
 E. Guzman-Herrera, A. Montiel, N. Breton, JCAP \textbf{11}, 002 (2024).


\bibitem{Kruglov:ab}
S. I. Kruglov, Mod. Phys. Lett. A. \textbf{35}, 2050291 (2020).

\bibitem{Novello:xy}
M. Novello, V. A. De Lorenci, J. M. Salim and R. Klippert, Phys. Rev. D. \textbf{61}, 045001 (2000).
\bibitem{Gralla:100}
S. E. Gralla, D. E. Holz and R. M. Wald, Phys. Rev. D \textbf{100},  024018 (2019).

\bibitem{Gralla:24018}
S.E. Gralla, D.E. Holz, R.M. Wald, Phys. Rev. D
\textbf{100}, 024018 (2019).

\bibitem{Feng:5103}
J. Peng, M. Guo, X.H. Feng, Chin. Phys. C \textbf{45}, 085103 (2021). 

\bibitem{Perlick:947}
V. Perlick, O.Y. Tsupko, Phys. Rep. \textbf{947}, 1 (2022).

\bibitem{Qiao:106}
C. K. Qiao,  Phys. Rev. D \textbf{106}, 084060 (2022).

\bibitem{Chena:512}
J. Chen, J. Yang, Eur. Phys. J. C  \textbf{85}, 512 (2025).


\bibitem{W110}
R. K. Walia,Phys. Rev. D \textbf{110}, 064058 (2024).

\bibitem{Okyay:009}
M. Okyay, A. \"Ovg\"un, JCAP \textbf{01}, 009 (2022).


\bibitem{Ahmedov:2016}
A. Abdujabbarov, M. Amir, B. Ahmedov, S. G. Ghosh, Phys. Rev. D \textbf{93}, 104004  (2016).


\bibitem{Kumar:100}
R. Kumar, S. G. Ghosh, A. Wang, Phys. Rev. D \textbf{100}, 124024 (2019).


\bibitem{Hawking:1975}
S. W. Hawking, Commun. Math. Phys. \textbf{43}, 199 (1975).

\bibitem{Sanchez:a} 
N. G. Sanchez, Phys. Rev. D \textbf{18}, 1030 (1978). 


\bibitem{Decanini:a}
Y. Decanini, G. Esposito-Farese, A. Folacci, Phys. Rev. D \textbf{83}, 044032 (2011).

\bibitem{Wei:2013}
S. W. Wei, Y. X. Liu, J. Cosmol. Astropart. Phys. \textbf{11}, 063 (2013).

\bibitem{Gibbons:235}
G. W. Gibbons and M. C. Werner, Class. Quantum Gravit.
\textbf{25}, 235009 (2008).

\bibitem{Gibbons:3047}
M. C. Werner, Gen. Relativ. Gravit. \textbf{44}, 3047 (2012).


\bibitem{Ishihara:abc} 
A. Ishihara, Y. Suzuki, T. Ono, T. Kitamura, H. Asada. Phys. Rev. D \textbf{94}, 084015 (2016).

\bibitem{Ishihara:def} 
A. Ishihara, Y. Suzuki, T. Ono and H. Asada, Phys. Rev. D \textbf{95}, 044017 (2017).


\bibitem{Kumaran:ao}
Y. Kumaran, A. \"Ovg\"un, Turk J Phys. \textbf{45}, 247 (2021).

\bibitem{Takizawa:101}
K. Takizawa, T. Ono, H. Asada, 	Phys. Rev. D \textbf{101}, 104032 (2020).

\bibitem{Jafarzade:kz}
 K. Jafarzade, Z. Bazyar, M. Jamil, Phys. Lett. B \textbf{864},  139390 (2025).

	\end{thebibliography}
\end{document}